\documentclass[twocolumn,prb,amsfonts,amssymb,amsmath,floats]{revtex4-2}

\usepackage{graphicx}
\usepackage{dcolumn}
\usepackage{multirow}

\usepackage[normalem]{ulem}
\def\expect#1{\left\langle#1\right\rangle}
\usepackage{braket}
\def\ie{i.e.\ }

\def\e{\varepsilon}

\def\w{\omega}
\def\e{\epsilon}

\def\k{\vec{k}}

\def\real{{\rm Re}}
\def\mat#1{\bm{#1}}
\usepackage{upgreek}
\usepackage{bbold}
\usepackage{bm}

\usepackage{xcolor}
\usepackage{placeins}

\begin{document}

\title{Strongly correlated multi-impurity models: The crossover from a single-impurity problem to lattice models}

\author{Fabian Eickhoff}
\author{Frithjof B. Anders}
\affiliation{Theoretische Physik 2, Technische Universit\"at Dortmund, 44221 Dortmund, Germany}

\date{\today}

\begin{abstract}
We present a mapping of various correlated multi-impurity Anderson models to 
a  cluster model  coupled to a number of  effective conduction bands 
capturing its essential low-energy physics. 
The major ingredient is the complex single-particle self energy matrix of the uncorrelated problem
that encodes the influence to the host conduction band
onto the dynamics of a set of correlated orbitals in a given geometry. While the real part of the self-energy matrix 
generates
an effective hopping between
the cluster orbitals, the imaginary part, or hybridization matrix, determines the coupling to the effective conduction electron bands in the mapped model.
The rank of the hybridization matrix determines the number of independent screening channels
of the problem, and allows the replacement of the
phenomenological exhaustion criterion by a rigorous mathematical statement.
This rank provides a distinction
between multi-impurity  models of the first kind and of the second kind.
For the latter, there are insufficient screening channels available, so that a singlet ground state must be driven by the inter-cluster spin
correlations. This classification provides a fundamental answer to the question of why ferromagnetic exchange interactions between local moments are irrelevant for the spin compensated ground state in dilute multi-impurity models, whereas the formation of large spins competes with the Kondo-scale in dense impurity arrays, without evoking a spin density wave.
The low-temperature physics of three examples taken from the literature are deduced from the analytic structure of the mapped model, demonstrating the potential power of this approach. 
Numerical renormalization group calculations are presented
for up to five site cluster. We investigate the appearance of frustration induced non-Fermi liquid fixed points
in the trimer,
and demonstrate the existence of several 
critical points of Kosterlitz-Thouless type at which 
ferromagnetic correlations suppress the screening of an additional effective spin-$1/2$ degree of freedom.

\end{abstract}

\maketitle

\section{Introduction}

The different competing phases in strongly correlated electron systems 
caused a lot of attention in the last 50 years. Heavy Fermions (HF) \cite{Grewe91,Maple95}
are a prominent example of a heavy Fermi liquid (FL) formation and superconducting phases \cite{Steglich79}.
Magnetically ordered phases \cite{LoehnysenWoelfeReview2007} can either develop out of a heavy FL with very low magnetic moments, 
or out of a local moment phase with almost unscreened magnetic moments \cite{Grewe91}.
Another prominent example
is the high-temperature superconductors, where a superconducting dome at finite doping is located next to 
an antiferromagnetically ordered Mott-Hubbard insulator \cite{HighTcReview2006}. External control parameters, such as doping or pressure, have been used to tune between phases of strongly correlated electron systems at low temperatures: strange-metals with non-Fermi liquid (NFL) properties have often been detected \cite{Maple95} in the vicinity of such a quantum critical point (QCP) \cite{Hertz76,Millis1993,LoehnysenWoelfeReview2007}. A sufficient understanding of such strange-metals, and their origin in strongly correlated electronic systems, is still lacking, and the underlying universality of strange-metal behavior that develops at a quantum critical phase transition is still subject of intense theoretical research.

The physics of the Heavy Fermions  is  governed by a competition between a heavy Fermi-liquid formation due to the Kondo effect \cite{Kondo1964}, and a magnetic ordering of localized spins due to the Ruderman-Kittel-Kasuya-Yosida (RKKY) interaction, both mediated by the light quasiparticles of the metallic host \cite{LoehnysenWoelfeReview2007}. Since these phases are orthogonal, the Doniach picture  \cite{Doniach77}
suggests that there exists a quantum phase transition between those two phases.

This scenario has triggered intensive work on the two-impurity Kondo problem as a simplified model aiming for
a microscopic understanding of a potential quantum phase transition (QPT) between a magnetically ordered and a heavy FL phase.
For the two-impurity Kondo problem \cite{Jones_et_al_1988}, however, it turned out that ferromagnetic exchange interactions between the local moments are irrelevant for the spin compensated ground state, and the antiferromagnetic QCP was shown to be unstable \cite{AffleckLudwigJones1995,Silva1996}: the two singlet fixed points are adiabatically connected by a continuous change of the conduction electron scattering phase.

The Doniach picture was already questioned  at the advent of early approaches to the periodic Anderson model (PAM) 
or the Kondo lattice model. Grewe \cite{GreweRKKY1988} pointed out, that this purely local picture 
neglects  the role of the band mediated interaction between the local moments. They
influence the Fermi liquid phase, as well as the formation of magnetically ordered phases
out of a heavy Fermi liquid phase, driven by the residual quasiparticle interactions. 
The theory of magnetism in such materials must include localized and itinerant magnetic order. 
Furthermore, Nozieres'  exhaustion scenario \cite{Nozieres1985,Nozieres1998}  challenged  the notion that the heavy Fermi liquid formation can be associated with individual Kondo effects at each lattice site in a periodic system, where the Kondo effect is mediated by the local density of states of the host conduction electrons. In spite of the criticism, the Doniach scenario remains a paradigm  \cite{LoehnysenWoelfeReview2007} even today  
in  illustrations of the potential origin of complex phase diagrams \cite{ColemanFrustration2010} in HFs.

While most of the HFs exhibit an antiferromagnetically (AF) ordered phase, there is a growing number of ferromagnetic HF compounds 
\cite{FMKL_Krellner2007,FMKL_Tran2014,FMKL_Szlawska2018,FMQCP_Okano2015,FMQCP_Yang2020,FMQCP_Kirkpatrick2020,FMQCP_Larrea2005,FMQCP_Khan2016,FMHF_Tada2016,FMHF_Brunning2008,FMHF_Rojas2012,FMQCP_Kotegawa2019,FMQCP_Shen2020,FMQCP_Steppke2013}.
Recent experiments on such ferromagnetic HF revealed strange-metal behavior when the Curie temperature is smoothly suppressed to zero via hydrostatic pressure \cite{FMQCP_Kotegawa2019,FMQCP_Shen2020} or chemical pressure \cite{FMQCP_Steppke2013}. Whereas quantum phase transitions in itinerant ferromagnets are always of first-order in the framework of Hertz-Millis-Moriya theory \cite{Hertz76,Millis1993}, experiments \cite{FMQCP_Kotegawa2019,FMQCP_Shen2020,FMQCP_Steppke2013} demonstrate the existence of local criticality with Kondo-destruction also in ferromagnetic HFs. This finding proves that the destruction of antiferromagnetism is not essential for the varied behaviors of strange-metals.
New theories for such ferromagnetic QCPs, aside from the first order spin density wave scenario, are, therefore, highly required and may provide new access in the context of strange-metals.
Impurity clusters of finite size, where the formation of a spin density wave is suppressed, hence, are a good starting point 
in order to obtain a microscopic understanding of why ferromagnetic correlations in larger correlated cluster can compete with the Kondo screening, even if such a competition 
has not been reported in
two- and three-impurity models \cite{Jones_et_al_1987,Eickhoff2018,PaulIngersent1996,Ingersent2005}.

The quest for a many-impurity problem that is solvable, and reveals interesting competing phases connected by a true QCP, triggered the investigation of the frustrated three impurity spin problems \cite{PaulIngersent1996,Ingersent2005,KoenigColemanKomijani2020,Woijcik2020,Kudasov2002,Savkin2005}. The connection to bulk materials, however, remains unclear, although it might be very helpful to illustrate the possibility of emerging complex phases. A key observation of these papers is the central role of magnetic frustration that is able to trigger more exotic phases in correlated materials.

In this paper, we present  an approach that is able to shed some light from a different perspective onto this old and fundamental question.  We start from the conventional multi-impurity Anderson model (MIAM) where $N_f$ correlated Hubbard atomic sites are hybridizing with Wannier orbitals of a single conduction electron band. This 
includes the well studied single-impurity Anderson and Kondo models \cite{BullaCostiPruschke2008}, and the periodic Anderson model (PAM), where $N_f$ is equal to the number of lattice sites of the host material $N_L$, as two opposite limits, as well as finite-size impurity clusters that become relevant for scanning tunneling microscopy (STM) or as toy models for magnetic frustration. We present a mapping for the original model to an effective low-energy MIAM  that is justified in the wide-band limit. The mapping accounts for
the conduction-band-mediated RKKY interaction and the delocalization of the correlated orbitals by effective hopping matrix elements between all orbitals, as well as the Kondo effect by the construction of effective band channels.

The number of effective screening conduction band channels in the mapped model depends on the lattice geometry and  the 
location of the impurities. The number of $k$-points on the Fermi surface of the host material provides the upper limit of the screening channels, challenging  the Doniach scenario of a Kondo screening of individual local spins by a single conduction band in the PAM. As a consequence of the mapping, the magnetic ordering, the heavy FL formation as well as the screening of the local moments are related to 
collective phenomena, involving a small number of  Kondo screening channels and the conduction band mediated effective interaction between the impurity orbitals.

In Refs.\ \cite{FMQCP_Komijani_2018,FMQCP_Shen2020} the paramagnetic-ferromagnetic transition in the Kondo lattice was studied within an independent bath approximation for each spin, such that the suppression of the Kondo temperature can be ascribed to the Kondo resonance narrowing in FM coupled single-impurity Kondo models \cite{Nevidomskyy_2009} in combination with an infinite number of coupled local moments.
However, due to the independent bath approximation, the exhaustion effect of the conduction electrons is neglected in such models.

Within the classification we present below, we propose a different mechanism leading to FM criticality in 
multi-impurity models, directly based on the reduced number of conduction band screening channels.
From the two- and three-impurity problem \cite{Jones_et_al_1987,Eickhoff2018,PaulIngersent1996,Ingersent2005} it is known that FM correlations do not compete with the single-ion Kondo effect and lead to a reduction of the Kondo temperature at most. However, if the number of impurities is large, such that there are not enough conduction screening channels
available,
the singlet ground state can not be interpreted in terms of the single-ion Kondo effect any longer, and additional collective mechanisms need to be taken into account.
We demonstrate that it is this collective singlet formation that competes with the formation of FM correlations between the local moments.

Since the delocalization and the collective screening of the individual local moments are both realized by operators 
responsible for the antiferromagnetic part of the RKKY interaction, ferromagnetic couplings lead to a competition between (localized) magnetic order and the (delocalized) heavy FLs. 
Consequently, the competition is rather between (delocalizing) AF and (localizing) FM RKKY interactions, than between (delocalizing) Kondo and (localizing) RKKY coupling, as usually assumed \cite{FMQCP_Komijani_2018,FMQCP_Shen2020}.
 
We demonstrate the formation of a ground state with finite magnetic moment and ferromagnetic spin-spin correlations between the local moments in multi-impurity models 
belonging to the class that adiabatically evolves to the PAM for a large number of correlated orbitals.
Due to the finite number of correlated orbitals a spin density wave scenario can be excluded in this case.
This ferromagnetic type of ground state is beyond the scope of the generic two impurity model where a spin singlet is always formed at sufficient low temperatures. The transition between a spin-singlet and a spin full ground state is accompanied by a QCP, at which ferromagnetic correlations lead to a suppression of the screening of an effective spin-$1/2$ moment, \ie, a linear combination of the local moments in real space. 
The QCP is stable against any kind of symmetry breaking and the transition can be driven by several parameters of the model.

Magnetic frustration arises when the nearest and next-nearest hopping matrix elements in the mapped model
become equally strong leading to competing antiferromagnetic interactions.
We study the impurity trimer in a $C_3$-symmetric setup, and review the frustration induced NFL fixed points \cite{PaulIngersent1996,Ingersent2005} within our effective low-energy model. While at intermediate strengths of AF RKKY interaction $K_\text{RKKY}>K_\text{RKKY}^{c,1}$ the NFL fixed point is stabilized, we establish the existence of an upper bound $K_\text{RKKY}^{c,2}$, at which the NFL fixed point gets unstable, and the system becomes a FL at low temperature. 
Since the RKKY interaction needs to dominate over the Kondo temperature $T_K$ in order to allow for magnetic frustration, the NFL fixed point completely disappears in the phase diagram, if the Kondo temperature exceeds this upper bound $T_K>K_\text{RKKY}^{c,2}$.

One of the strengths of our effective low-energy mapping is that it incorporates the FM and AF RKKY interaction as well as the generated potential scattering terms:
It naturally incorporates the correct  symmetries of the original  multi-impurity models coupled to only one single conduction band \cite{Eickhoff2018,AffleckLudwigJones1995,Silva1996}. We do not need to 
add artificial Heisenberg exchange couplings, which might lead to unphysical fixed points as known from the two impurity model \cite{Eickhoff2018,AffleckLudwigJones1995,Silva1996}, to 
realize and explore the competing phases.

The paper is organized as follows. After introducing the precise definition of our model in Sec.\ \ref{sec:model}, we provide a preliminary overview
of the results  in Sec.\ \ref{sec:prelim}.
While the mapped low-energy MIAM is derived in Sec.\ \ref{sec:low-energy-model}, 
we introduce the rank of the interaction matrix as quantitative classification in multi-impurity problems of the first and second kind
in Sec.\ \ref{seq:classification} and discuss several impurity-cluster configurations in different spatial dimensions. 
The strength of our mapping is demonstrated in Sec.\ \ref{sec:literature} where we revisit three different problems investigated in the literature using sophisticated  methods
and predict the central result of each problem:
(i) the ferromagnetic ground state of the dilute PAM \cite{Titvinidze2015,Potthoff2013} at half-filling,
(ii) the ferromagnetic ground state in the one-electron limit of the Kondo-lattice model \cite{KLM-Sigrist91}, as well as (iii) the
scaling of the critical $U_c$ of the Mott transition in the PAM with nearest-neighbor hybridization found in an elaborate dynamical mean field calculation \cite{HeldBulla2000}.
The limits of the mapping are addressed in Sec.\ \ref{sec:applicability}.
Our numerical renormalization group (NRG)  results on three, four and five impurity clusters are presented in Sec.\ \ref{sec:NRG-results}.   In Sec.\ \ref{sec:C3_frustration} we study the impurity trimer in a $C_3$-symmetric setup, and review the frustration induced NFL fixed points \cite{PaulIngersent1996,Ingersent2005}.
For short 1d impurity chains in Sec.\ \ref{sec:MIAM_1d}, we report on a sequence of
Kosterlitz{\textendash}Thouless-type phase transitions as a function of the host band filling and the strength of Coulomb interaction.
In Sec.\ \ref{sec:KT-QPT} we study dense impurity arrays, such that the fixed point evolves from a singlet at half-filling to a maximally polarized multiplet at the band edge, and we discuss and explore the role of magnetic frustration at intermediate band fillings in Sec.\ \ref{sec:kinks-magnetic-frustration}.
For dilute multi-impurity models in Sec.\ \ref{sec:KT-QPT_FM} the situation is vice versa, i.e. starting from a maximally polarized multiplet at half-filling we can drive the system across several QCPs to a spin-singlet ground state. We conclude the paper
with a short summary and an outlook in Sec.\ \ref{sec:summary}.

\section{Theory}
\label{sec:theory}

\subsection{Model}
\label{sec:model}

Although strongly correlated electron systems have a large number of incarnations in particular when
applied to realistic material science, we focus on the most elementary version in this paper
that targets impurity clusters on surfaces as well as the elementary modeling of HF.
These models can be easily generalized to more complex situations if needed,
for instance to multiple correlated   $3d$-orbitals as required in transition  metal ions, 
but show already rich physics that is worth presenting
from a different perspective.

Quantum impurity systems are typically embedded in a metallic host which is represented by
a non-interacting tight-binding model
\begin{align}
\label{eq:band}
 H_{\rm host}=-\sum_{ i,j \sigma}t_{ij} c^\dagger_{i,\sigma}c_{j,\sigma} 
 = \sum_{\k\sigma} \e_{\k\sigma}  c^\dagger_{\k \sigma}c_{\k\sigma} 
\end{align}
that is diagonalized in $k$-space in a periodic lattice.
$i,j$ label all lattice points $\vec{R}_i\in S_L$
where $S_L$ defines the set of all lattice points. The dimension of $S_L$ is $N_L$. $\e_{\k\sigma}$
denotes the band dispersion obtained from Fourier transformation of the matrix $t_{ij}$,
and can also include a Zeeman term due to an external magnetic field not considered in this paper. 
$\e_{\k\sigma}$ becomes a continuous function of $\k$
for $N_L\to \infty$.  The diagonal element $t_{ii}$ accounts for the
local orbital energy and is used to shift the band center of the conduction band.
In this paper, we restrict ourselves to nearest neighbor tight-binding models for keeping the parameter space simple, but
our approach is applicable to arbitrary dispersions $\e_{\k\sigma}$.

The $N_f$ impurities 
are   located at the positions $\vec{R}_l \in  S_f$ and are modeled by an atomic  Hubbard Hamiltonian
\begin{align}
\label{eqn:himp}
H_{\text{corr}} &= \sum_{l,\sigma}\epsilon^f_{l\sigma} f^{\dagger}_{l,\sigma}f_{l,\sigma}
+\frac{1}{2}\sum_{l,\sigma}U_l f^{\dagger}_{l,\sigma}f_{l, \sigma}f^\dagger_{l,\bar{\sigma}}f_{l,\bar{\sigma}},
\end{align}
where $f_{l}^{(\dagger)}$ destroys (creates) an electron in the single-impurity orbital at site $l$. The 
on-site energies are labeled by $\e^f_l$,  $\bar\sigma=-\sigma$, and $U_l$ denotes the on-site Coulomb repulsion.
In general,  correlated $3d$ or $4f$-shells contain many more degrees of freedom. Here, we focus on the essentials to
keep the number of free parameters to a minimum. We have situations in mind where crystal electric fields 
separate the ground state doublet energetically from higher excitations and spin-orbit coupling between the conduction electrons and the local degrees of freedom can be neglected. However, the mapping and the classification introduced below are applicable to an arbitrary number of orbital degrees of freedom as well.
The mapping introduced below is applicable to arbitrary locations, but throughout the paper we
focus on a finite dimensional subset $S_f \subset S_L$ of the underlying lattice.

Since we are only considering the thermodynamic equilibrium, we can either explicitly use a chemical potential $\mu$
to adjust the different fillings, or we absorb $\mu$ by collectively shifting all single particle energies, $t_{ii}$ and $\e_l^f$, by
the same amount and leave $\mu=0$. We adapt the latter convention and investigate the effect of different conduction band
fillings by shifting the band center  $\e_c=t_{ii}$.

The most general coupling between the two orthogonal subsystems is given by the spin-diagonal hybridization term
\begin{eqnarray}
\label{eqn:hybrid}
 H_{\rm hyb}&=&\sum_{l,m\sigma}V_{m,l} c^\dagger_{m \sigma}f_{l\sigma}+{\rm h.c.}
\\
&=& 
 \sum_{\k,l\sigma}V_{\k,l} c^\dagger_{\k \sigma}f_{l\sigma}+\rm h.c.
\nonumber
\end{eqnarray}
where $V_{\k,l}$ is obtained by a Fourier transformation,
\begin{eqnarray}
V_{\k,l}  &=&  \frac{1}{\sqrt{N_L}} \sum_m V_{m,l} e^{-i\k \vec{R}_m} .
\end{eqnarray}

In this paper, we only consider  a local hybridization, i.~e.\ $V_{m,l} = \delta_{ml} V_l$,
$V_{\k,l} =   V_l \exp(-i\k \vec{R}_l)/\sqrt{N_L}$, and
a nearest neighbor hybridization $V_{m,l} =V$ for $\vec{R}_m$ and $\vec{R}_l$  being
nearest neighbor sites and $V_{m,l} =0$ otherwise, corresponding to
$V_{\k,l}=  -(V/t) (\e_{\k}
-\e_c)\exp(-i \vec{k}\vec{R}_l)
$ \cite{HeldBulla2000}.
The strength of the coupling is typically discussed in terms of
$\Gamma_{0,l,_\sigma}=\pi V^2_l\rho_\sigma(0)$, which describes the effective hybridization of a single-impurity with
a conduction band density of states (DOS) $\rho_\sigma(\e)$. 
Although $\rho_\sigma(\e)$ can be spin-dependent, we consider only spin-independent host DOSs
throughout this paper.

The total Hamiltonian of the system  is given by
\begin{eqnarray}
H &=&  H_{\rm host} + H_{\text{corr}}  +  H_{\rm hyb}.
\end{eqnarray}
This formulation includes two  well established and well understood limits.
If $S_f=S_L$, and $N_f=N_L\to \infty$, we recover the PAM. If
$S_f$ only contains a  single site, the model is known as the single-impurity Anderson
model that was accurately solved using the NRG  \cite{KrishWilWilson80a,KrishWilWilson80b} and the Bethe ansatz  \cite{AndreiFuruyaLowenstein83,Schlottmann89}
 almost 40 years ago.
If the number of sites $1<N_f \ll N_L$ is small and finite,
we refer to a multi-impurity Anderson model (MIAM)
whose simplest realization is the two-impurity Anderson model (TIAM)
\cite{Jones_et_al_1987,Eickhoff2018}. 
Multi-band versions of the model have been addressed using an interleave approach to the 
construction of Wilson chains \cite{MitchellGalpinBulla2014} with its virtues and limitations.

\subsection{Preliminaries}
\label{sec:prelim}

\begin{figure}[t]
\begin{center}
\includegraphics[width=0.48\textwidth]{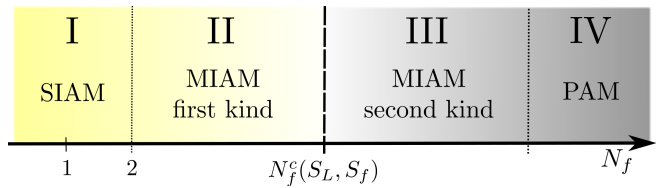}

\caption{Schematic partitioning of the models in four different categories: I denotes the limit of a single-impurity 
problem, section II the MIAMs of the first kind, section III the MIAMs of the second kind
and section IV the MIAM with infinite number of correlated lattice sides recovering the periodic Anderson model.}
\label{sec:categories-MIAMs}

\end{center}
\end{figure}

At the heart of this paper lies the extension of the low-energy mapping developed for the two-impurity model \cite{Eickhoff2018}
to the  multi-impurity situation ($N_f>2$)  and the consequences that can be concluded 
from  this mapping.
The single-impurity problem \cite{Wilson75} as well as the two-impurity problem \cite{Jayaprakash1981,Jones_et_al_1987,AffleckLudwigJones1995,Silva1996} have been extensively investigated 
over the last four decades and are well understood.

The interest for the two-impurity problem originates in the Doniach scenario \cite{Doniach77} for Heavy Fermions  (HF) \cite{Grewe91}
which relates the origin of the magnetic ordering found  in some of the HFs
to the competition between the single-ion  Kondo effect \cite{Kondo1964}
screening the local moments and the RKKY interaction \cite{RKKY1954,lit:Kasuya1956,lit:Yosida1957,Florian2017,Lichtenstein2000,Zhou2010,Allerdt2015,Aristov1997,Masrour2016,Nefedev2014} favoring magnetic ordering of those moments.  Although lacking a rigorous mathematical
proof, such appealing visualizations of the complex physics in  HF \cite{LoehnysenWoelfeReview2007,ColemanFrustration2010,ColemanLecture} are popular even today
\cite{ColemanLecture}, since they provide a simple picture that can  intuitively be grasped. 
This picture was already challenged by  Nozieres'  exhaustion scenario \cite{Nozieres1985,Nozieres1998}, as well as
by the observation that HF magnetism can form even out of a heavy FL phase but with strongly reduced magnetic moments \cite{GreweRKKY1988,Grewe91}. This indicates that the  Doniach scenario is too simplistic and does not reflect the full physics in such complex correlated electron systems.

Figure \ref{sec:categories-MIAMs} summarizes the four categories of the MIAM 
which we can  mathematically rigorously  distinguish within our mapping presented below.
The two well established limits  of the MIAM model, the single-impurity Anderson model  ($N_f=1$) 
and the periodic Anderson model ($N_f=N_L$) are located at the opposite end of the figure, where the horizontal axis 
denotes the number of impurities $N_f$.

We define the MIAM of the first kind by our ability to map the low-energy physics problem onto 
an effective coupled multi-impurity cluster that couples the $N_f$ localized orbitals  to $N_b=N_f$  effective conduction bands: 
The number of effective conduction bands, $N_b$, exactly matches 
the number of impurities which allows for spin-singlet formation by a compensated multi-channel  Kondo effect \cite{CoxZawa98}.

The first example of such a MIAM of the first kind  is the well understood  two impurity model \cite{Jayaprakash1981,Jones_et_al_1987,Jones_et_al_1988,IJW92,AffleckLudwigJones1995,Silva1996}, $N_f=2$. 
Jones and Varma  showed \cite{Jones_et_al_1987} that the model can be 
mapped onto a two-impurity, two-band model in the even/odd parity basis. 
The induced orbital hopping  \cite{Eickhoff2018} is responsible for the antiferromagnetic (AF)  exchange,
the asymmetry of the couplings to the two bands results in the ferromagnetic (FM) part of the RKKY interaction  
\cite{Jones_et_al_1987}, while the two bands allow  screening the impurity moments via a two-stage Kondo effect.
The QCP that emerges, if the energy dependence of the hybridization functions in the even/odd basis is neglected \cite{Jones_et_al_1988}, is just a consequence of unphysical approximations \cite{AffleckLudwigJones1995,Silva1996} which automatically restores a special kind of particle-hole symmetry that is absent in the original model.
Since the antiferromagnetic RKKY interaction, which is required to drive the phase transition, in the full model is dynamically generated from the same contributions that break this special symmetry, the QCP is replaced by a continuous crossover once the full energy dependence is correctly incorporated.
Any approximative solution of multi-impurity problems, as a toy model with regard to quantum criticality in HFs therefore, needs to ensure the absence of a QCP in the two impurity limit.
Other examples of the MIAM of the first kind are trimer models \cite{PaulIngersent1996,Ingersent2005,KoenigColemanKomijani2020} involving three effective conduction bands.

Depending on the details of the lattice topology and the geometric arrangement of the impurities,
we find a critical value $N_f^c(S_L,S_f)$ above which the MIAM maps onto a low-energy
multi-impurity cluster  that couples to a reduced number of effective conduction bands $N_b<N_f$. 
We call these types of problems the MIAM of the second kind,
indicated  by the category III in  Fig.\  \ref{sec:categories-MIAMs}. This reduced number of coupled conduction bands
has profound consequences for the magnetic properties of the system:
a large local moment, that is formed at low temperatures,
cannot be completely screened by the multi-channel Kondo effect. We argue below that the periodic Anderson model (PAM)
is a particular example for such an MIAM problem of the second kind: The screening of the local moments must involve the antiferromagnetic RKKY-induced inter site exchange coupling, which competes with the ferromagnetic ones. As we demonstrate below, this competition results in several QCPs in multi-impurity models of the second kind.

Nozieres \cite{Nozieres1985,Nozieres1998} and others \cite{TahvildarJarPruFre1999} 
already suggested that only the  fraction $T_K\rho(0)$  of the conduction electrons can contribute to the Kondo screening in the PAM, 
$T_K$ being the single-impurity Kondo temperature and
$\rho(0)$ being the conduction electron density of states at the chemical potential.
Therefore, there are not enough conduction electrons available for ensuring the Kondo screening of all  local $f$-moments by independent Kondo screening mechanisms. 

The singlet ground state formation 
in a heavy Fermi liquid must be based  on a different mechanism than simply
extending the single-impurity Kondo effect to periodic structures. Although the PAM is mapped 
onto an effective single-site problem \cite{Kuramoto85,Grewe87,Kim90,Jarrell95,PruschkeBullaJarrell2000} embedded into
a lattice self-consistency condition in the context of  the dynamical mean field theory (DMFT) \cite{Georges96},
indicating a simple connection between the single-ion Kondo effect and the Kondo lattice problem,
Pruschke and collaborators \cite{PruschkeBullaJarrell2000}  interpreted the occurring chemical potential dependent reduction of the effective conduction electron density of states in the effective single-site problem  in terms of  Nozieres' exhaustion
scenario. Moreover, Hollender and Bulla demonstrated in Ref.\ \cite{Hollender2012} a striking difference between $T_K$ and the low energy scale $T_c$ of the PAM: Using a constant DOS in their DMFT calculations they found a strong dependence of $T_c$ on the filling of the conduction band, whereas $T_K$ remains nearly constant in this case.

Based on our low-energy mapping presented below, we  provide a different perspective on 
the heavy FL formation in the PAM and Kondo lattice model (KLM). It replaces 
the phenomenological exhaustion scenario with a rigorous mathematical criterion and connects the local Kondo screening
and magnetic ordering within the DMFT approach to the mechanism  \cite{PruschkeHM1990} known from the Hubbard model.

\subsection{Low-energy effective multi-impurity model}
\label{sec:low-energy-model}

Since an exact solution of complex multi-impurity correlated electron systems is not known in most 
of the interesting cases, when the number of impurities exceeds $N_f>2$, 
the challenge is to find an appropriate approximation to nevertheless extract the relevant low-temperature physics.
We propose a mapping onto an effective low-energy model 
which can be used to analyze the emergence of 
free local moments in a variety of different situations, and allows us to understand their screening as well as the potential magnetic ordering.  
Throughout the rest of the paper, we assume the absence of a complicated magnetically ordered phase
in the host as well as on the correlated sites. We focus on problems in which the $z$-component of the spin is a good quantum number.

The effect of the host conduction band onto the dynamics of the correlated lattice sites
is determined by  the spin-diagonal hybridization function matrix \cite{JabbenGreweSchmitt2012},
\begin{align}
 \Delta_{lm,\sigma}(z)=\frac{1}{N_c}\sum_{\k}\frac{V^*_{\k, l}V_{\k,m} e^{i\k(\vec{R}_l-\vec{R}_m)}}{z-\e_{\k\sigma}},
 \label{eq:delta-mat-def}
\end{align}
derived from Eq.\ \eqref{eqn:hybrid}

The exact real space multi-impurity Green's function matrix 
of the dimension $2N_f\times 2N_f$ in spin-orbital space is reduced to
two $N_f\times N_f$ block matrices for  spin-diagonal problems 
and is  given by the matrix
\begin{eqnarray}
\mat{G}_\sigma(z) &=&[ z - \mat{E}_\sigma - \mat{ \Delta}_\sigma (z)]^{-1},
\end{eqnarray}
in the absence of the Coulomb interaction, $U_l=0$.
The matrix $\mat{E}_\sigma$ is diagonal and contains the single-particle 
energies of the localized orbitals, $\e_{l\sigma}$,
\footnote{Note that for the translational invariant periodic Anderson model (PAM)
the Green's function matrix is diagonalized by Fourier transformation where $\Delta(\k) = V^2/(z-\e_{\k})$.},
and the matrix elements of the self-energy matrix $\mat{ \Delta} (z)$ are given in Eq.\ 
\eqref{eq:delta-mat-def}. 
For $U_l>0$, the self-energy matrix $\mat{ \Delta} (z)$ is augmented by a correlation contribution
$\mat{\Sigma^U}_\sigma(z)$, $\mat{ \Delta} (z) \to\mat{ \Delta} (z) +\mat{\Sigma^U}_\sigma(z) $
which was the starting point of the screening channel analysis in Ref.\ \cite{MitchellBulla2015}.

The energy dependence of $\Delta_{lm,\sigma}(z)$ can be neglected in the wide band limit, $V_{l}/D\rightarrow 0$. Its effect onto the local impurity dynamics is mainly determined by the complex matrix elements $\Delta_{lm,\sigma}(- i0^+)$ for a  $\rho_\sigma(\w)$ that is almost constant on the relevant low-energy window.
We also have to be careful with the distance dependency of the off-diagonal matrix elements $\Delta_{lm,\sigma}(z)$. The definition \eqref{eq:delta-mat-def} reveals that the larger the distance,
the more pronounced the frequency oscillation of  $\Delta_{lm,\sigma}(z)$ is close to Fermi energy.
The error of the approximation is estimated by the first derivative of the imaginary part 
of $\Delta_{lm,\sigma}(z)$, $d\Gamma_{lm,\sigma}(\w)/d\w$ \cite{Eickhoff2018}. While the oscillations are very pronounced for an isotropic dispersion of the conduction electrons in 1d,
resulting in significant corrections in 1d, the derivative becomes $R$ independent in 3d -- see Eq.\ (61) in Ref.\ \cite{Eickhoff2018}. The proposed approximation is valid in the regime $d\Gamma_{lm,\sigma}(\w)/d\w\to 0$ which can always be ensured for $V_{l}/D\rightarrow 0$.
In Sec.\ \ref{sec:applicability} we discuss the applicability and limitations of the mapping in detail.


We divide the complex matrix element $\Delta_{lm,\sigma}(-i0^+)$ 
into its real and imaginary part: $\Delta_{lm,\sigma}(-i0^+) = {\rm Re} \Delta_{lm,\sigma}(-i0^+) + i\Gamma_{lm,\sigma}$.
We absorb the effective inter-orbital hopping matrix elements ${\rm Re} \Delta_{lm,\sigma}(-i0^+)$
into the energy matrix $\mat{E}_\sigma\to \mat{E}_\sigma + {\rm Re} \mat{\Delta}_\sigma(-i0^+)$.
If we are only interested in the  dynamics of the multi-impurity cluster degrees of freedom, 
the problem can be mapped onto an effective problem
where the charge fluctuation matrix $\Gamma_{lm,\sigma}$ is generated by a fictions set of conduction bands.
For that purpose, we diagonalize the Hermitian matrix $\Gamma_{lm,\sigma}$,
\begin{eqnarray}
\mat{\Gamma}_\sigma &=&  \mat{U}^h_\sigma \mat{\Gamma^d}_\sigma \mat{U}_\sigma 
\end{eqnarray}
where the $N_f$ eigenvalues $\Gamma^d_{n\sigma} = \pi \bar V_{n\sigma}^2 \rho^0_\sigma(0)$ are interpreted as coupling of the new orbital $n$ to 
the n-th new effective conduction band with the same DOS $\rho^0_\sigma(\e)$ as the original conduction band DOS  at half filling, $\e_c=0$,
and the hybridization strength is given by $\bar V_{n\sigma}$. 
Note that a spin-dependent hybridization function, for instance generated by an external magnetic field, leads
to different effective orbitals for the two spin orientations. Although we focus on spin-independent hybridization functions in this paper, we have numerically implemented the full spin-full approach in our NRG code.

This is justified since we are only interested in the different fixed-point structure of the model
and not in the accurate calculation of the low-temperature crossover scale. Its precise number
is also determined by the high-energy degrees of freedom \cite{Wilson75,BullaCostiPruschke2008,AU-PTCDA-monomer}.
Then the Green's function is approximated by
\begin{eqnarray}
\mat{G}_\sigma(z)&\approx& \mat{U}_\sigma [ z - \mat{E'}_\sigma - i\mat{ \Gamma}^d]^{-1}_\sigma \mat{U} ^h_\sigma,
\end{eqnarray}
where the energy matrix $ \mat{E'}$, 
\begin{eqnarray}
 \mat{E'}_\sigma &=& \mat{U}_\sigma [  \mat{E}_\sigma + {\rm Re} \mat{\Delta}_\sigma(-i0^+)]  \mat{U} ^h_\sigma
\end{eqnarray}
contains diagonal and hopping terms between all correlated impurity orbitals in the new eigenbase diagonalizing $\mat{\Gamma}_\sigma$.
The approximation is limited to a range of frequencies $z$ for which $\Delta_{lm,\sigma}(z)\approx \Delta_{lm,\sigma}(si0^+)$ with $s={\rm sign}({\rm Im} z)$.

Consequently, the same low-frequency single-particle Green's function matrix is  generated by the  
effective single-particle Hamiltonian
\begin{eqnarray}
\label{eqn:MIAM_hyb}
H'_{\rm sp} &=& H_{\rm cl} + H_{\rm deloc}
\end{eqnarray}
in the new eigenbase of $\mat{\Gamma}_\sigma$ in the limit $V_l/D\to 0$. 
The cluster part of the mapped  Hamiltonian $H'_{\rm sp}$,
\begin{eqnarray}
\label{eq:H-cl}
H_{\rm cl}&=& \sum_{m,l}  E'_{lm\sigma} f^\dagger_{l\sigma}f_{m\sigma},
\end{eqnarray}
defines the single-particle Hamiltonian of the correlated orbitals in the new basis that
have acquired additional orbital hopping terms mediated by the conduction band of the host.
The second part,
\begin{eqnarray}
\label{eq:H-deloc}
 H_{\rm deloc} &=& \sum_{n=1}^{N_f}
 \sum_{\k\sigma}
 (\e_{\k\sigma}-\e_c)c^\dagger_{\k,n,\sigma}c_{\k,n,\sigma}
\\
&& \nonumber
 +  
 \sum_{n=1}^{N_f}  \sum_{\k\sigma} \left(\frac{\bar V_{n\sigma}}{\sqrt{N_c}} c^\dagger_{\k,n,\sigma}f_{n\sigma} +h.c.\right),
\end{eqnarray}
includes the new $N_f$ effective conduction band degrees and the flavor diagonal
coupling to the cluster orbitals for each conduction band flavor $n$. Note, however, that
the total number of particles in each flavor $n$ is in general not  conserved, since this operator does not commute with the single-particle cluster Hamiltonian $H_{\rm cl}$.

Whereas the hybridizations $\bar{V}_{n\sigma}$ of the mapped model are exclusively determined by the Fermi-surface, all high energy conduction band states contribute to the dynamics of the cluster $H_\text{cl}$
via the energy matrix elements $ E'_{lm\sigma}$.
The real part of the complex hybridization function, and the effective hopping elements $t^\text{eff}_{lm\sigma}$, respectively, can be deduced via a Hilbert transformation 
\begin{eqnarray}
\label{eq:ReDelta_Hilbert}
t^\text{eff}_{lm\sigma}={\rm Re}\Delta_{lm,\sigma}(-i0^+)=\frac{1}{\pi}\int_{-\infty}^{\infty}d\e \frac{\Gamma_{lm,\sigma}(\e)}{\e}
\end{eqnarray}
that incorporates some information on the \textit{whole} energy dependence of the coupling functions.
These hopping elements generate  the antiferromagnetic part of the RKKY interaction 
and simultaneously lead to destruction of the QCP in the two-impurity limit \cite{AffleckLudwigJones1995,Silva1996,Eickhoff2018}
since it is a relevant perturbation of the fixed point \cite{AffleckLudwigJones1995}.

The correlated MIAM is recovered after the 
local Coulomb matrix elements in $H_{\text{corr}}$, Eq.\ \eqref{eqn:himp}, has also been rotated into
the new orbital basis as well, and added to the single-particle Hamiltonian $H'_{\rm sp}$.
This leads to a complicated, coupled multi-impurity problem that still contains
the  full spatial correlations in contrast to a local approximation in real space that is employed by 
the DMFT.

This mapping generates $N_f$ fictitious conduction bands
labeled with the index $n$ and the corresponding, orthogonal single-particle orbitals, augmented by 
new orbital energies and an inter-orbital hopping, both included in the matrix elements $E'_{lm\sigma}$. 
As a side product of this mapping, we have separated the AF part of the RKKY interaction that is generated by the
inter-site matrix elements of  $E'_{lm\sigma}$ from the Kondo screening channels.
Since the hybridization strengths $\Gamma_{n\sigma}$ are in general all different, multi-stage screening of local moments is found in such situations.
Furthermore, the differences generate the FM part of the RKKY interaction \cite{Jones_et_al_1987,Eickhoff2018}.
In addition we found that for large $N_f$, only a few $\Gamma_{n\sigma}$ are different from zero, therefore, the number of Kondo screening channels is typically much smaller than $N_f$.

This mapping was previously investigated \cite{Eickhoff2018} in 
the two-impurity Anderson model ($N_f=2$) 
where the two conduction bands represent states with even and with odd parity.
In this case, one can either use the full energy dependency of the even-parity and 
the odd-parity band \cite{AffleckLudwigJones1995,Silva1996,LechtenbergAnders2014,Lechtenberg2018} in an numerical renormalization group (NRG) 
\cite{BullaCostiPruschke2008} calculation, or investigate the mapped Hamiltonian \eqref{eqn:MIAM_hyb}
with a particle-hole symmetric band density of states.
Both Hamiltonians, the original MIAM, as well as the mapped Hamiltonian, produced the same RG fixed points for a featureless and spin-independent initial
$\rho(\w)$, and the same spin-spin correlation functions in the wide band limit, establishing the quality of the mapping for $N_f=2$.
It was shown that 
the particle-hole asymmetry in the even and odd conduction band dynamically generates an effective hopping between
the two local orbitals, which is responsible for the AF part of the RKKY interaction via an exchange mechanism once $U_l>0$.
The FM part of the RKKY interaction is generated by the imbalance between the two eigenvalues of $\mat{\Gamma}_\sigma$.

\subsection{Classification of the  multi-impurity problem by the ${\rm rank}(\mat{\Gamma}_\sigma)$}
\label{seq:classification}

Having   an effective low-energy multi-impurity model at hand,
we can now rigorously define different classes of MIAMs, as well
as distinguish between a multi-impurity model of the first kind versus one of the second
kind.  Mathematically, the hybridization matrix $\mat{\Gamma}_\sigma$ has exactly $N_f$ eigenvalues 
if its rank is equal to the number of impurity orbitals, i.~e.\  ${\rm rank}(\mat{\Gamma}_\sigma)=N_f$. 
Since the original model only contains a single conduction electron band,
the phase correlations between different lattice sites encoded in  Eq.\ \eqref{eq:delta-mat-def}
are responsible for the fact that the rank of the matrix is often less than the  number of impurity orbitals: ${\rm rank}(\mat{\Gamma}_\sigma)< N_f$. 
Therefore, we use the  ${\rm rank}(\mat{\Gamma})$ to classify the MIAM into two categories:
A multi-impurity problem of  the first kind requires that ${\rm rank}(\mat{\Gamma}_\sigma)=N_f$, while ${\rm rank}(\mat{\Gamma}_\sigma)<N_f$
defines a multi-impurity problem of the second kind.

Throughout the rest of the paper, we only consider a paramagnetic host with spin-degenerated bands in the
absence of an external magnetic field.
Although, we maintain the spin index in $\mat{\Gamma}_\sigma$ for consistency, both $\Gamma$-matrices are the identical and therefore, have the same rank. It is up to a future study to investigate problems where
the matrices for the different spin orientations might have different ranks.

The multi-impurity problem of the first kind is an example of a compensated multi-channel Kondo problem \cite{CoxZawa98}:
there are always enough conduction band channels available for
a complete screening of all local moments via a multi-stage Kondo effect \cite{CoxZawa98}. 
When reducing the temperature of the system, the details of the
eigenvalues $\Gamma^d_{n\sigma}$ of $\mat{\Gamma}_\sigma$ define a cascade of low-energy scales at which the local moments are quenched by $1/2$ until the singlet ground state is reached.
Although  the transfer matrix $E'_{lm\sigma}$ is responsible 
for generating an effective low-energy Heisenberg model, representing the AF part of the RKKY interaction, the pre-quenching of the moments via the effective Heisenberg couplings 
is not needed to obtain a singlet ground state.
Consequently, ferromagnetic exchange couplings between the local moments are irrelevant with respect to the singlet ground state for models of the first kind.
However, interesting physics can arises in problems that contain magnetically frustrated systems
requiring at least three impurities \cite{PaulIngersent1996,Ingersent2005,KoenigColemanKomijani2020,Kudasov2002,Savkin2005}. Our mapping provides an ideal tool
to investigate which physical condition the original model must fulfill in order to reach  the critical parameter regimes reported for the trimer Kondo models \cite{PaulIngersent1996,Ingersent2005,KoenigColemanKomijani2020,Woijcik2020,Kudasov2002,Savkin2005}
or their Anderson model incarnations \cite{MitchellLogan2009,MitchellLogan2010}

The two-impurity Anderson  (TIAM) or Kondo model is
a typical representative of a multi-impurity problem of the first kind where always a singlet ground state is generated - with the
exception of peculiar geometric conditions \cite{AU-PTCDA-dimer,LechtenbergEickhoffAnders2017,Eickhoff2018}
where ${\rm rank}(\mat{\Gamma}_\sigma)=1$ is found.  Although the interest in the TIAM was driven
by the Doniach scenario, the originally reported quantum phase transition \cite{Jones_et_al_1987}
between a two-stage Kondo singlet and an RKKY induced singlet turned out to be  an artifact of the approximation. This model shows a crossover between both phases accompanied by a continuous variation of the scattering phase \cite{AffleckLudwigJones1995,Silva1996} which is also included in our
mapped model \cite{Eickhoff2018}. The QPT is destroyed by the real part of 
$\mat{\Delta}_\sigma(-i\delta)$ inducing a hopping term in the cluster that is a relevant perturbation
in the vicinity of the QPT  \cite{AffleckLudwigJones1995,Silva1996,Eickhoff2018}.

The TIAM is an ideal  system to explicitly understand the origin of the rank reduction in our effective model,
since  the original mapping by Jones and Varma always leads to a coupling to two conduction bands. 
Following the arguments of Ref.\ \cite{LechtenbergEickhoffAnders2017}, or inspecting the imaginary part of $\Delta_{lm,\sigma}(z)$,
Eq.\ \eqref{eq:delta-mat-def}, in the even or odd parity basis
for certain dispersions $\e_{\k}$ and relative distances $\vec{R}_l-\vec{R}_m$ between the impurities, yields a vanishing
of the energy-dependent coupling function $\Gamma_\sigma(\w)$  at the chemical potential of a power-law form 
$|\w|^\alpha$, where $\alpha>1$. For such an exponent of  pseudo-gap coupling functions
the local moment fixed point has been proven to be stable \cite{WithoffFradkin1990,pgSIAM_Ingerset1998,Vojta2006} in the RG flow. Therefore, the approximation made in the effective low-energy model \eqref{eqn:MIAM_hyb}, by neglecting the full energy dependency of the bands, is fully justified since the low-energy fixed point remains unaltered. 
The rank of the coupling function matrix $\mat{\Gamma}_\sigma$ is  a simple measure to identify the number of independent effective conduction electron channels that can be potentially used for the screening of local moments by the Kondo effect.

A interesting consequence arises for large $N_f$, for instance in  the PAM where $N_f=N_L$. Let us consider a  very large
but finite system with periodic boundary conditions. In this case, we know that the problem can be diagonalized in 
$\k$ space: The new multi-impurity orbitals are labeled also by the quantum number $\k$, and acquire a very complicated 
non-local Coulomb matrix. The single-particle matrix $\mat{\Delta}_\sigma(z)$, however, must be  diagonal in $\k$, and the matrix 
elements take the very simple form
\begin{eqnarray}
\Delta_{\k\sigma}(z) &=& \frac{|V_{\k}|^2}{z-\e_{\k\sigma}},
\end{eqnarray}
which is the well known self-energy of the  $f$-lattice Green's function.
As a consequence, only the $\k$-values for which $\e_{\k\sigma}=0$ holds, yield a finite $\Gamma_{\k\sigma}$ in the mapped model.
Therefore,  ${\rm rank}(\mat{\Gamma}_\sigma)\ll N_f$ for the PAM, and the number of available screening channels is related to the size of the Fermi surface and not
 the number of correlated orbitals. 
We can conclude that in one-dimension  ${\rm rank}(\mat{\Gamma}_\sigma)\le 2$ since the Fermi surface is discrete
and only contains two points.

Our concept of classifying the MIAM including the periodic model in terms of  ${\rm rank}(\mat{\Gamma}_\sigma)$
allows a much more precise definition of the phenomenological exhaustion principle: 
for a Kondo screening in the MIAM there are only ${\rm rank}(\mat{\Gamma}_\sigma)$ screening channels available.
Obviously, this definition is only governed by the single-particle properties, introduced by the arrangement of the impurities,
the underlying lattice, and the host dispersion $\e_{\k\sigma}$.  This mathematically precise definition, however, is able to replace the phenomenological
notion of a fraction $\rho_\sigma(0) T_K$ of electrons contributing to the Kondo screening, which requires the definition of $T_K$
although $T_K$ became a questionable quantity in MIAM.

This finding is a strong indicator that the singlet ground state in the PAM is caused by a different mechanism: it is
driven by the hopping matrix elements $E'_{lm}$ delocalizing the local impurity electrons within the $f$-impurity subsystem and not by $N_f$ independent conduction electron channels,
as already conjectured by Grewe \cite{GreweRKKY1988} more then 30 years ago. 
For this second kind of MIAMs, the formation of large spins due to ferromagnetic exchange couplings competes with the self-screening of the correlated electrons, and leads to several QCPs in the phase diagram of such models.
Interesting physics also arises from the competition between
self-screening of the correlated impurity cluster and magnetically frustration due to long range  hopping matrix elements $E'_{lm\sigma}$ in finite dimensions. By inspecting Eq.~\eqref{eq:delta-mat-def}, one can conclude that $E'_{lm\sigma}$ decays rather rapidly in higher spatial
dimensions destroying the physics of magnetic frustration in the limit $d\to\infty$, in accordance with
the arguments of Metzner and Vollhardt \cite{MetznerVollhardt89}, Brandt and Mielsch \cite{BrandtMielsch89}
as well as M\"uller-Hartmann \cite{MuellerHartmann89}.

The question of the number of screening channels in a multi-impurity model was also raised in Ref.\ \cite{MitchellBulla2015}. The authors focused on the full energy dependence of $\mat{\Gamma}(\w)$ 
which only allows a definite 
statement in the limit of $|\vec{R}_l-\vec{R}_m|\to 0$ for all impurity combinations $l,m$ or for a particular 
high-symmetry
point, where one or several $\Gamma(\w)=0$ over the full frequency range.
A general construction of multi-band NRG chains can be found for instance in the Supplemental Material
of Ref.\ \cite{OpenChains2017}.
The strength of our approach, however, lies in the revelation of the low-energy physics of the model
even in complicated setups when the details of the high-energy physics only influence the 
crossover scales but not the different emerging low-energy fixed points. 
This is  achieved  by focusing   on the low-energy description in the vicinity of the chemical potential in the spirit of Wilson's original ideals \cite{Wilson75}, and this approximation becomes exact
in the wide band limit.

\subsubsection{The ${\rm rank}(\mat{\Gamma}_\sigma)$ for finite impurity-cluster in various dimensions}
\label{sec:Rank2d3d}

In the PAM the number of k-points on the Fermi surface of the host material provides the upper limit for the number the available screening channels, independently of the structure and the dimension of the underlying lattice. Since the number of decoupled $f$-orbitals in the PAM must continuously develop out of the MIAM with a finite number $N_f$ of correlated impurities, we study the reduction of ${\rm rank}(\mat{\Gamma}_\sigma)$ for finite impurity clusters with different geometries and in different dimensions. For this purpose we consider a simple cubic lattice with nearest neighbor hopping $t$ such that the dispersion $\epsilon_{\vec{k}}$ in $d$ dimensions reads $\epsilon_{\k\sigma}=-2t\sum_{i}^d\cos(k_i a) + \e_c$. In the following we concentrate on dense impurity arrays where all the impurities are placed next to each other. Dilute impurity configurations can always be deduced from a dense array, by shifting the on-site energy $\epsilon^f_i$ of the depleted sites to infinity. Therefore, the ${\rm rank}(\mat{\Gamma}_\sigma)$ of the dense array serves as an upper limit for any depleted configuration that can be deduced from the dense case.

In one dimension, the Fermi surface consists of two single points which determine the rank of the charge-fluctuation matrix for the PAM: ${\rm rank}(\mat{\Gamma}_\sigma)=2$. Consequently for any finite number $N_f$ of correlated impurities we can conclude ${\rm rank}(\mat{\Gamma}_\sigma)\leq2$ and $N_f^c=3$. Every MIAM in 1d with $N_f\geq N_f^c$ belongs to the MIAM of the second kind and thus exhibits QCPs due to FM correlations between the local moments in its parameter space.

\begin{figure}[t]
\begin{center}
\includegraphics[width=0.4\textwidth]{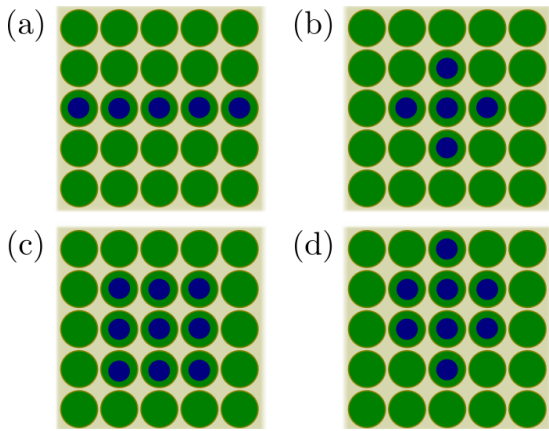}

\caption{MIAM with 2d simple cubic lattice, containing uncorrelated lattice orbitals (green) and correlated impurities (blue). The individual panels depict different geometries of the impurity configuration leading to a different number of decoupled orbitals $N_\text{free}=N_f-{\rm rank}(\mat{\Gamma}_\sigma)$. (a) $N_\text{free}=0$, (b) $N_\text{free}=1$, (c) $N_\text{free}=1$ and (d) $N_\text{free}=2$. 
}
\label{fig:RankGamma_2d}
\end{center}
\end{figure}

In higher dimensions the Fermi surface itself becomes a continuum in the thermodynamic limit and an argumentation analog to the 1d case is not possible.
Hence we focus on some explicit configurations in 2d which are schematically depicted 
in Fig.\ \ref{fig:RankGamma_2d}.

If the impurities are placed in line along the $x$-direction as schematically depicted in Fig.\ \ref{fig:RankGamma_2d} (a), we can diagonalize the charge-fluctuation matrix in the limit $N_f\to\infty$ via a 1d Fourier transformation
\begin{align}
\Gamma_{k_x}\propto\sum_{k_y}|V_{\k}|^2\delta(\epsilon_{\k}).
\label{eq:Gamma_kx_2d}
\end{align}
For a half-filled conduction band, $\epsilon_c=0$, one can always find a $k_y$ such that $\k=(k_x,k_y)^T$ belongs to the Fermi surface and, consequently, $\Gamma_{k_x}\not=0$ which implies a MIAM of the first kind: ${\rm rank}(\mat{\Gamma}_\sigma)=N_f$.
Deviations from half-filling, $\e_c\not=0$, lead to a small number of $k_x$-points for which no $k_y$ can be found such that $\k=(k_x,k_y)^T$ belongs to the Fermi surface. 
In this case the MIAM with $N_f\to\infty$ belongs to the second kind, however, the number of decoupled orbitals $N_\text{free}=N_f-{\rm rank}(\mat{\Gamma}_\sigma)$ remains small and for $N_f\leq 50$ a numerical evaluation yields ${\rm rank}(\mat{\Gamma}_\sigma)=N_f$ for various fillings.

For a finite number of $N_f=5$ impurities, which are arranged as depicted in panel 
Fig.\ \ref{fig:RankGamma_2d}(b) we can analytically calculate ${\rm rank}(\mat{\Gamma}_\sigma)$ using the irreducible representation of the $C_4$ point group. In this basis one obtains three 1d subspaces with $\Gamma_{n\sigma}\not=0$, for a general filling $\e_c$, and one 2d subspace.
The 2d subspace contains the correlated orbital in the center of the impurity array $f_c$ and the even combination $f_e=\frac{1}{2}\sum_i f_{i,o}$, where $f_{i,o}$ denotes the annihilation operator of the outer impurities. The charge-fluctuation matrix of this 2d subspace for arbitrary fillings $\e_c$ reads
\begin{align}
\mat{\Gamma}_{\text{even},\sigma}=
\begin{pmatrix}
  \Gamma^\text{outer}_e & \Gamma^\text{o/c}_e \\
  \Gamma^\text{o/c}_e &  \Gamma^\text{center}_e     
\end{pmatrix}
=\Gamma_0
\begin{pmatrix}
  (\e_c/2t)^2 & \e_c/2t \\
  \e_c/2t & 1       
\end{pmatrix},
\end{align}
and exhibits an incomplete rank for every $\e_c$ due to $\text{det}(\mat{\Gamma}_{\text{even},\sigma})=0$. We can conclude ${\rm rank}(\mat{\Gamma}_\sigma)=4=N_f-1$ for the configuration in Fig.\ \ref{fig:RankGamma_2d}(b), which, consequently, belongs to the second kind of MIAMs.

For arbitrary arrangements of $N_f$ impurities on a 2d simple cubic lattice, the charge-fluctuation matrix 
$\mat{\Gamma}_\sigma$ can be numerically evaluated using Eq.\ \eqref{eq:delta-mat-def}. For the configuration in panel (c) of Fig.\ \ref{fig:RankGamma_2d}, for instance, we obtain ${\rm rank}(\mat{\Gamma}_\sigma)=8=N_f-1$, whereas the evaluation of ${\rm rank}(\mat{\Gamma}_\sigma)$, for an arrangement as depicted in Fig.\ \ref{fig:RankGamma_2d} (d), yields ${\rm rank}(\mat{\Gamma}_\sigma)=6=N_f-2$. 
The analytical and numerical evaluation of the rank of the charge-fluctuation matrix $\mat{\Gamma}_\sigma$ for a simple cubic lattice in 2d, as well as in 3d,  indicates, that the number of available screening-channels is proportional to the number of the outermost impurities of a certain arrangement. This finding is compatible with the fact, that ${\rm rank}(\mat{\Gamma}_\sigma)$ for the PAM is limited to the number of $\vec{k}$-points on the Fermi surface of the host material, which in general is proportional to $N^{d-1}$.

For STM experiments, the impurity cluster on a 2d surface of a 3d crystal is of particular interest. In order to qualitatively study such situations we used the exact 2d surface Greens function 
$G^0_{2d,\sigma}(\vec{k}_{||},\w)$, with $\vec{k}_{||}=(k_x,k_y)^T$, of a semi-infinite 3d simple cubic lattice, which can be found in \cite{STM_Mitchell2015,STM_Derry2015}, to construct the complex hybridization matrix 
$\mat{\Delta}_\sigma(z)$.
If the number $N_f$ of impurities is small, we found a MIAM of the first kind in general. Configurations which belong to the second kind of MIAMs in the pure 2d case, however, exhibit a clear hierarchy of hybridizations $\Gamma_{n\sigma}$ in the mapped model. Hence, due to the finite temperature in experiments, the fully spin-compensated ground state may not be reached since the smallest Kondo temperature is exponentially suppressed.
For a large number of impurities arranged in a dense cluster one will continuously reach the limit of a fully covered 2d surface. In that case we can diagonalize the charge-fluctuation matrix $\mat{\Gamma}_\sigma$ via a 2d Fourier transformation which, according to \cite{STM_Mitchell2015,STM_Derry2015}, yields
\begin{align}
\label{eq:Gamma_2d_surface}
\Gamma_{\k_{||}\sigma}\propto \begin{cases}
\sqrt{1-\left( \e_{\k_{||}}^\text{2d}/2t\right)^2} & \text{if}\quad \left|\e_{\k_{||}}^\text{2d}/2t\right| \leq 1\\
\quad\quad\quad 0 & \text{else} \end{cases},
\end{align}
and, consequently, reveals a MIAM of the second kind.

\subsection{Constructing effective cluster models in the local moment regime}

Having an effective multi-impurity model at hand, we can use the results of Sec.\ \ref{sec:low-energy-model}
to propose a two step process in order to gain some physical insight in the low-energy properties of the original model.
In a first step, we set the couplings $\bar V_n$ to the effective conduction bands to zero, and focus on the decoupled cluster dynamics. After understanding the ground state and the elementary excitations within the cluster, we couple the cluster to the neglected conduction bands. Such a procedure 
implies a certain hierarchy of energy scales: the Coulomb repulsion, being the largest energy scale causes a local moment formation, dividing
the cluster Hilbert space in irreducible subspaces of the total spin. The  hopping matrix defines the intermediate energy scale
selecting the ground state multiplet of the cluster. In the last step, the effective local moment fixed point of the decoupled cluster becomes
unstable due to the coupling to the neglected conduction bands.  

For this energy hierarchy one can employ a two step Schrieffer-Wolff type \cite{SchriefferWol66} transformation. In the first step, such a transformation is applied to the decoupled cluster. This leads to a finite size $t-J$ model for large $N_f$ as used in the context of the high temperature superconductors \cite{HighTcReview2006}. Depending on the particle-hole asymmetry, a pure spin model might emerge, favoring locally antiferromagnetically aligned spins that might order for $N_f\to \infty$, or a more complicated model with two and three site interactions. In a second Schrieffer-Wolff type \cite{SchriefferWol66} transformation, an effective Kondo coupling is obtained  between the ground state multiplets and the now included  
coupling to the previously neglected conduction bands.

In the SIAM \cite{KrishWilWilson80a,KrishWilWilson80b,BullaCostiPruschke2008}, the system flows to  
the same strong-coupling fixed point in the case of a dominating coupling to conduction electrons but without a clear signature of the local moment fixed point. Therefore, the  energy hierarchy outlined above is not essential 
in a full NRG and only helps shaping our physical intuition in
some limited cases. Hence, our low-energy MIAM also contains the correct physics for the cases in which the hybridization strengths $\bar \Gamma_{n\sigma}$ dominate over the magnetic exchange terms that would be generated by the first Schrieffer-Wolff-type transformation.
In this case the local moments are starting to be partially screened before the interaction between the magnetic moments becomes relevant.

\subsection{Applying the mapping to some cases discussed in the literature} 
\label{sec:literature}

\begin{figure}[t]
\begin{center}
\includegraphics[width=0.45\textwidth]{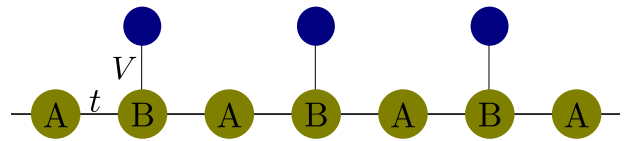}

\caption{The diluted periodic Anderson model in 1d: the local impurities are only connected to the B sublattice.
}
\label{fig:2:depleted-PAM}
\end{center}
\end{figure}

\subsubsection{The dilute Kondo lattice model in 1d}
\label{sec:dilutePAM}

An interesting version of a diluted Kondo lattice model in 1d was investigated by Potthoff and 
collaborators \cite{Titvinidze2015,Potthoff2013} at half filling. 
The correlated electron sites are only coupled to the B-sublattice
of the bi-partite lattice as  depicted in Fig.\ \ref{fig:2:depleted-PAM}. The authors report that the ground state is ferromagnetic, and the total spin $S=(N_f-1)/2$ contains one local moment less than the total number of correlated sites. 

This surprising finding can be very easily explained by employing our mapping onto a low-energy multi-impurity model.
The 1d chain \cite{Titvinidze2015,Potthoff2013} with a constant lattice spacing $a$ forms a bipartite lattice.
A simple nearest-neighbor tight-binding band structure, $\e_k=-2t\cos(ka)$, was considered  by  Potthoff and collaborators \cite{Titvinidze2015,Potthoff2014,Potthoff2015,Potthoff2013}. 
Starting from a half-filled ($k_\text{F}=\pm\pi/2a$) 
conduction band, the real and imaginary parts
of $\Delta_{lm,\sigma}(-i0^+)$ are given by
\begin{subequations}
\label{eqn:1d-Delta}
\begin{align}
{\rm Im}\Delta_{lm,\sigma}(-i0^+)&=\pi V_l V_m \rho_\text{1d}(0) \cos\left(\frac{\pi}{2} \frac{R_{lm}}{a}\right)\\
{\rm Re}\Delta_{lm,\sigma}(-i0^+)&=V_l V_m \int_{-D}^{D}d\e \frac{\rho_\text{1d}(\e)\cos\left(\cos^{-1}\left(\frac{\e}{D}\right)\frac{R_{lm}}{a}\right )}{\e}
\label{eq:17b}
\end{align}
\end{subequations}
where $R_{lm}=R_l-R_m=a(l-m)$ denotes the relative distance between the two impurities at $R_l$ and $R_m$.
If the impurities are placed on the same (different) sublattice, i.e., $R_{lm}=2na$ ($R_{lm}=[2n+1]a$) with $n\in\mathbb{Z}$,
the real (imaginary) part of $\Delta_{lm}(-i0^+)$ vanishes. For $V_l=V$ we obtain
\begin{align}
\label{eq:1d-lattice-Delta}
\Delta_{lm,\sigma}(-i0^+)=\begin{cases} \pm i\Gamma_0  &\text{same sublattice} \\
{\rm Re}\Delta_{lm,\sigma}(-i0^+)\quad &\text{different sublattice} \end{cases}
\end{align}
where $\Gamma_0 =\pi V^2 \rho_\text{1d}(0)$.
Then, the  hybridization function matrix of the correlated lattice sites at the sublattice B, as depicted in Fig.\ \ref{fig:2:depleted-PAM},
is given by a purely imaginary spin-independent
matrix
\begin{align}
 \bm{\Delta}_\sigma=i\Gamma_0
\begin{pmatrix}
  +1 & -1 & +1 & \cdots \\
  -1 & +1 & -1 & \cdots \\
  +1 & -1 & +1 & \cdots \\
  \vdots & \vdots &\vdots &\ddots          
\end{pmatrix}
\end{align} 
at zero frequency.
The matrix is finite dimensional for a finite lattice of size $N_c=2N_f$ with
 ${\rm rank}(-i\mat{\Delta}_\sigma)=1$, such that only a single eigenvalue $\bar \Gamma_{0}=N_f\Gamma_0$ is different from zero.
Only one of the $N_f$ new orbitals couples to an effective conduction electron band, and in real space
a  $1/N_f$ fraction of the local moments on each lattice site is screened by the flow to the strong coupling fixed point.
The other $N_f-1$ orthogonal orbitals are disconnected from a conduction band since  ${\rm Re}[\mat{\Delta}_\sigma]=0$. Since the asymmetry of the coupling constants is responsible for the FM part of the RKKY interaction \cite{Jones_et_al_1987,Eickhoff2018}, those remaining $N_f-1$ effective moments are aligned ferromagnetically at low temperatures, as reported using a density-matrix renormalization group (DMRG) calculation \cite{Titvinidze2015,Potthoff2014,Potthoff2015,Potthoff2013}.  Therefore, our mapping provides a simple and intuitive understanding of such a complex correlated system as the dilute PAM in 1d.

The question arises whether this finding for the 1d diluted PAM is symmetry related or 
a consequence of the phase relation in Eq.\ \eqref{eq:delta-mat-def},
connecting two lattice sites via a single conduction band.
Let us imagine the case in which the coupling to the impurity site at the origin is much larger than
to all others: $V_0\gg V_l, l\not = 0$. Then we can solve the problem in two steps: solve a single-impurity problem by setting $V_l=0, l\not = 0$, and once the low-energy fixed point of that problem is reached, switch on the other couplings $V_l$.
The diagonal component of the conduction electron Green's function in real space,
\begin{align}
G_{ ii,\sigma}(z)=G^0_{ ii,\sigma}(z)+V^2_0G^0_{ i0,\sigma}(z)G_{\text{imp},\sigma}^U(z)G^0_{0i,\sigma}(z),
\end{align}
is obtained from the exact equation of motion and determines the  local density of states prior to switching on $V_l$.
$G^0_{ i j}(z)$ denotes the free propagator from $R_i$ to $R_j$ 
\begin{align}
 G^0_{ ij,\sigma}(z)&=\frac{1}{N}\sum_{k}\frac{e^{ik(R_{i}-R_{j})}}{z-\epsilon_k}\nonumber\\
&=\int_{-D}^D \frac{\rho^0_{\rm 1d}(\omega)\cos\left\{\frac{R_i-R_j}{a}\cos^{-1}\left[\frac{\omega}{D}\right]\right\}}{z-\epsilon}
\end{align}
and $V^2_0 G_{\text{imp},\sigma}^U(z)$ the t-matrix of the impurity located at the origin, calculated for a finite $U$.  For a particle-hole symmetric band, the real part of $ G^0_{ij,\sigma}(z)$ essentially vanishes for $(R_i-R_j)=2na$, and $n\in \mathbb{Z}$, at low frequencies.
If we substitute  
a Lorentzian approximation for $G_{\text{imp}}^{U=0}(z)$ (Kondo effect), we obtain the approximate solution
\begin{align}
\label{eq:rho-eom-disance-R}
&\rho_{R_{i},\sigma}(\omega)\approx\tilde{\rho}_{R_{i},\sigma}(\omega)=\quad\rho_\text{1d}^0(\omega)\,\,-\nonumber\\
&\left[\rho_\text{1d}^0(\omega)\cos\left\{R_{i}\cos^{-1}\left[\frac{\omega}{D}\right]\right\}\right]^2
\frac{\pi^2V^4_0\rho_\text{1d}^0(0)}{\omega^2+\left[\pi V^2_0 \rho_\text{1d}^0(0)\right]^2}
\end{align}
for the local conduction electron spectral function at site $R_i$.
A careful analysis of this DOS reveals  a pseudo-gap formation of the spectrum at all other correlated impurity sites: 
the larger the distance, the faster the DOS oscillations in energy space, the smaller the energy interval
of the pseudo-gap. Since the pseudo-gap always vanishes quadratically, $\Gamma_{i,\sigma}(\w) \propto |\w|^2$ in this energy  window,  a local magnetic moment of another impurity coupled to the conduction electrons at site $R_i$  remains unscreened in the limit $T\to 0$.  

As a consequence, the artificial  distortion of the coupling constants $V_l$, in favor of a single dominating hybridization, provides a real-space interpretation of our finding that ${\rm rank}(\mat{\Gamma}_\sigma)=1$ in the 1d dilute PAM.  The delocalized orbital coupling to the single effective band for $V_l=V_0$ is 
adiabatically deformed to the localized orbital at the origin for $V_l\ll V_0 (l\not =0)$,
which does not alter the rank of the matrix. In our approach, the real space nature of this orbital is encoded in the eigenvector of $\mat{\Gamma}_\sigma$ corresponding to the single finite eigenvalue.

In an approach that includes the full energy dependency of the conduction bands in an effective $N_f$ band model, we obtain $N_f-1$ channels
that decouple quadratically at the chemical potential, and, therefore, cannot screen a local moment \cite{WithoffFradkin1990,pgSIAM_Ingerset1998,Vojta2006}: The rank is independent of coupling asymmetries and governed by the dispersion $\e_k$ and the topology of the impurity locations.
Consequently, the fixed point of the full model is correctly captured by the effective low-energy Hamiltonian, even for large $V/D$ apart from the wide band limit, contrary to local quantities.
As demonstrated by DMRG calculations \cite{Titvinidze2015,Potthoff2013}, the size of the magnetic moment of the ground state is independent from the strength of $V/D$. However, whereas it possesses mainly $f$-character in the wide-band limit $V/D\to 0$, it continuously shifts into the host, and becomes a pure conduction band quantity in the limit $V/D\to\infty$.
This flow of the magnetic moment, from $f$- to $c$-character, obviously is absent in the effective low-energy description for the wide-band limit.
\subsubsection{Limit of the one-electron Kondo-lattice model}
\label{sec:dilute_model}

Extending the consideration to the full PAM in 1d at half-filling,
the problem can be interpreted as coupling the two dilute PAM problems
of each sublattice via the hopping matrix elements defined in Eq.\ \eqref{eq:17b}.
On each sublattice, one spin is screened while the remaining $N_f/2-1$ spins are 
ferromagnetically aligned. The mapping \eqref{eq:1d-lattice-Delta} 
generates an effective hopping between the orbitals
on the different sublattices. This yields a band formation within the correlated sites 
as well as an AF coupling between the ferromagnetically
aligned spins on the sublattices favoring an antiferromagnetic ordering.

Sigrist et al.\ \cite{KLM-Sigrist91} derived rigorous results for the ground states for the one-electron Kondo lattice model
almost 30 years ago. The authors have proven that the ground state is of incomplete ferromagnetic order with 
$S_{\rm tot} = (N_f -1)/2$ for antiferromagnetic Kondo couplings always present in the PAM as well. 
We demonstrate that is can also be concluded from our mapped model.  For that purpose we recall that we have to shift the chemical potential to the lower band edge, i.\ e.\ $\mu\to -2td$ or $\e_c\to 2td$,
while keeping the correlated site singly occupied, where $d$ is the dimension of the simple cubic lattice. 
For $|\k_F|\to 0$, we obtain $\Gamma_{lm,\sigma}=\Gamma_0$
and as a consequence again ${\rm rank}(\mat{\Gamma}_\sigma)=1$. The same holds for $\k_F\to \pi(1,\cdots,1)^T$. Therefore, one effective moment is screened, and
our mapping predicts a FM ground state of $S_{\rm tot} = (N_f-1)/2$ as proven by Sigrist et al.\ \cite{KLM-Sigrist91}.
Note that the limit $J\to \infty$ of the Kondo lattice model, where a ground state formed by local singlets is expected, is a singular point \cite{KLM-Sigrist91} and not accessible within the PAM. While the exact form of
the ground state and the ground state energy  still require 
detailed calculations, the rank of $\mat{\Gamma}_\sigma$ provides already a basic understanding of
the elementary magnetic properties of the model in the limit of a vanishing band filling.

\subsubsection{Mott transition in the periodic Anderson model with nearest neighbor hybridization}

As a third example we consider the  periodic Anderson model with nearest neighbor hybridization \cite{HeldBulla2000}.
The real-space matrix $\mat{\Delta}_\sigma(z)$ becomes diagonal in $\k$-space,
\begin{eqnarray}
\Delta_{\k\sigma}(z) &=&  \frac{|V|^2}{t^2} \frac{(\e_{\k}-\e_c)^2}{z-\e_{\k}}.
\end{eqnarray}
Clearly, the imaginary part of $\Delta_{\k\sigma}(\w-i\delta)$  vanishes for a half-filled conduction band, $\e_c=0$,
for all $\k$, reducing ${\rm rank}(\mat{\Gamma}_\sigma)=0$: no
effective conduction band couples to the impurity orbitals, and one is left with 
a decoupled infinite size cluster of $f$-orbitals.

The real part of $\Delta_{\k\sigma}(z)$, takes the form
\begin{eqnarray}
\tilde t_{\k} &=&  -\frac{|V|^2}{t^2} \e_{\k},
\end{eqnarray}
leading to a renormalized band dispersion for $U=0$ 
\begin{eqnarray}
\e_{\k}^{f} = \e^f -\frac{|V|^2}{t^2} \e_{\k}
\end{eqnarray}
in $\k$-space. Our approximation maps the PAM with nearest neighbor hybridization onto
an effective single-band Hubbard model. 
The band dispersion  of this decoupled $f$-electron subsystem
is identical to those of the host conduction electron but rescaled by the factor $(V/t)^2$
and a different sign. 

Without any further calculations, we can conclude that this model must undergo 
a Mott-Hubbard metal-insulator  transition at two different critical values $U^f_{c1},U^f_{c2}$ for $T\to 0 $
depending on whether one comes from the metallic or the insulating site of the QPT \cite{Georges96,Bulla1999}.
The values for $U^f_{c}$ are
related to the established results for the Hubbard model by the scaling factor
$V^2/t^2$. This prediction of our mapping
perfectly agrees with the elaborate DMFT calculation by Held and Bulla \cite{HeldBulla2000}.  
They found  (i)  a Mott-Hubbard metal-insulator  transition in their full DMFT calculation
and (ii) a  scaling of $U^f_{c} = (V/t)^2 U_c^{\rm Hub}$
for small $V$. The scaling factor is modified for larger hybridization strength $V$, 
exactly as discussed in Sec.\ \ref{sec:low-energy-model}: corrections must be included for the deviations from the wide-band limit. This demonstrates 
the power and the potential of our mapping for the qualitative understanding
of such correlated lattice models.

\subsection{Discussion: Applicability and limitations of the mapping}
\label{sec:applicability}

In this section we address the possible shortcomings of our mapping neglecting the full energy dependency of 
$\mat{\Delta}(\w-i0^+)$.
The static properties of the mapped model based on the approximation $\mat{\Delta}(\w \pm i0^+)\approx\mat{\Delta}(\pm i0^+)$ become exact if the hybridization function matrix $\mat{\Delta}(z)$ is featureless on the scale of $T_K$ around the Fermi energy. This is always fulfilled in the wide-band limit $V/D\to0$ since then $T_K/D\to0$ even if $T_K/\Gamma_0$ is held constant ($V\propto\sqrt{D}$).

In the case of a MIAM of the first kind the frequency dependence of $\mat{\Delta}(z)$ only leads to a modification of the Kondo and RKKY scale but no additional fixed points will occur.
In contrast to that one has to be a bit more careful when studying a MIAM of the second kind, where some of the effective conduction band channels $l$ decouple due to  $\Gamma_l=0$. 
If we take into account the whole energy dependence, the imaginary part of the hybridization matrix will be non-diagonal for $\w\not=0$ in general. 
We can not completely exclude some subtile reentrance of the screening due to the remaining off-diagonal matrix elements
which only vanish for $\w=0$ and couple the high energy degrees of freedom of the individual Wilson chains. 
At least for the variety of quite different models discussed in Sec.\ \ref{sec:literature} 
and treated with various methods that do not rely on any mapping and include the full energy dependence, however, we know that the finite frequency off-diagonal matrix elements do not alter the fixed point structure of the ground state, even beyond the wide band limit.

Concentrating on the remaining energy dependence of the diagonal elements, the vanishing eigenvalues at $\w=0$ indicate a pseudogap, $\Gamma(\w)\propto |\w|^\alpha$, with exponent $\alpha$ and width $\delta$ in the corresponding channel of the full model.
For $\alpha<1$ or large particle-hole (PH) asymmetry, screening of local moments may become possible at a critical coupling strength $V_c/D$ \cite{WithoffFradkin1990,pgSIAM_Ingerset1998,Vojta2006} and additional fixed points can emerge beyond the wide-band limit that are not included in the mapped model.

Even if the local moment fixed point is stable, we nevertheless need to consider the case of a 
width $\delta\ll T_K$:
Around $T\approx T_K$ the screening of the $f$-local moment by conduction band electrons 
sets in when lowering the temperature further,
and the system flows to an unstable intermediate strong coupling fixed point. Just on the scale of $\delta$ the remaining Wilson chain sites gradually decouple and contribute to the free local moment of the ground state, which, consequently, possesses mainly $c$-character \cite{LechtenbergEickhoffAnders2017}.
This unstable strong coupling fixed point and the redistribution of the location of the free local moment with increasing $T_K/\delta$ is not included in the mapped model, where the impurity is completely decoupled from the conduction band as is the case for $T_K\ll\delta$.
However, the stable fixed point is independent from the ratio of $T_K/D$, or $T_K/\delta$ respectively,
in contrast to a local moment coupled to a superconducting host with gap of width $\Delta_{sc}$
\cite{SC-NRG-I-1992,SC-NRG-II-1993,Hecht_2008,PradhanFransson2020}: 
Since the host consists of a hard-gap instead of a pseudogap, the Wilson chain is finite such that the RG stops at the scale of $\Delta_{sc}$ and the intermediate fixed point becomes stable for $T_K>\Delta_{sc}$.
The mapped model fails in such a situation and is only applicable for $T_K<\Delta_{sc}$.

The character and locational flow of the local moment, from $f$- to $c$-character when increasing $T_K/D$, or $T_K/\delta$ respectively, has been observed in the context of the dilute PAM, discussed in Sec.\ \ref{sec:dilutePAM}, and is also relevant for the dilute limit in 1d: 
For two impurities placed on the same sublattice of a half-filled chain, the mapped model contains only one single conduction band due to ${\rm rank}(\mat{\Gamma})=1$, and the dilute limit seems to be absent. However, a detailed analysis of the energy dependence of the decoupled channel \cite{LechtenbergEickhoffAnders2017} reveals $\alpha=2$ but $\delta\to0$ for $|R_i-R_j|\to \infty$, such that $T_K/\delta\to\infty$ for a constant $V/D$ and both impurities get screened independently at $T_K$. The crossover to the stable local moment (LM) fixed point occurs on the scale $\delta\propto|R_i-R_j|^{-1}$ \cite{LechtenbergEickhoffAnders2017}. Hence, in a potential experimental realization of such a 1d TIAM at a fixed temperature $T>0$, one would observe a crossover to the dilute limit when the distance-dependent energy gap width $\delta(|R_i-R_j|)\approx T$.  
In higher dimensions the crossover to the dilute limit at finite temperature is well included in the mapped model since the Fermi surface is a continuum. 
Starting from the dense limit with ${\rm rank}(\mat{\Gamma})<N_f$ one 
will find ${\rm rank}(\mat{\Gamma})=N_f$ once all the impurities are separated by at least
some  characteristic minimal
distance.

\section{Numerical renormalization group results for multi-impurity problems }
\label{sec:NRG-results}

\subsection{Emerging trimer with three conduction electron bands}
\label{sec:C3_frustration}

In the last decades the role of magnetic frustration for HFs  has been discussed in the context of strongly correlated multi-impurity models \cite{LoehnysenWoelfeReview2007,KoenigColemanKomijani2020,Ingersent2005,PaulIngersent1996,Woijcik2020,Kudasov2002,Savkin2005}. 
Since it was realized that the two-impurity problem
lacks  complexity and the  reported QCP appears to be an artifact of the approximation \cite{AffleckLudwigJones1995,Silva1996}
made in the originally approach, nowadays the focus lies on models with three local moments coupled to an arbitrary number of conduction bands \cite{Ingersent2005,PaulIngersent1996,KoenigColemanKomijani2020,Woijcik2020,Woijcik2020,Kudasov2002,Savkin2005}. 
These papers in particular focus on the emergence of a frustration induced NFL fixed point that
might be related to QCPs found in bulk materials. In this section we 
review these kinds of models from the perspective of our effective low-energy mapping.

In the case of three identical correlated orbitals, the structure of the effective low-energy model reduces to four free parameters:
The diagonal elements of $\mat{\Delta}_\sigma(-i0^+)$ are equal for all impurities, whereas the off-diagonal elements $\Delta^\text{off}_{ij,\sigma}(-i0^+)$ depend on the geometric arrangement of the impurities and the structure, as well as the filling, of the underlying lattice of the crystal. For $\Gamma^\text{off}_{ij,\sigma}=\text{Im}\Delta^\text{off}_{ij,\sigma}(-i0^+)=0$ each of the correlated orbitals couples to its own, independent conduction band, as studied in \cite{KoenigColemanKomijani2020}. In general, however, the impurities are coupled to the same conduction band, which implies $\Gamma^\text{off}_{ij,\sigma}\not=0$, such that a precursive diagonalization of $\mat{\Gamma}_\sigma$ is necessary in order to obtain an ''independent-bath'' description of the model.

For a generic setup of the three impurity problem in a paramagnetic 
metallic host, the Hamiltonian does not preserve the symmetry of the $C_{3}$ point group, and, consequently, at least either $\Gamma^\text{off}_{12}\not=\Gamma^\text{off}_{23}$ or $\text{Re}\Delta^\text{off}_{12}(-i0^+)=t^\text{eff}_{12}\not=t^\text{eff}_{23}$. 
Note that we dropped the spin index $\sigma$ for clarity since the parameters are spin-independent.
For such a situation a frustration-induced NFL fixed point is unlikely to be found, since either the Kondo-scale or the RKKY-scale induces an imbalance that prevents a possible frustration. 

In some special setups, however, the Hamiltonian may preserve the symmetry of the $C_{3}$ point group,  
as studied in 
Refs.\ \cite{PaulIngersent1996,Ingersent2005}, where the Kondo-impurities are placed in a crystal with an isotropic dispersion relation in an arrangement of an equilateral triangle with impurity separation $\Delta R$. 
In this case, the authors found three different fixed points.

If the ferromagnetic RKKY interaction dominates, the local moment that forms at intermediate temperature is screened by the three independent effective conduction bands, leading to a FL with spin-singlet ground state. This corresponds to the trivial case of individual Kondo-screening of the three local moments for $\Delta R\to\infty$ and is referred to as the independent Kondo fixed point (FP).
In the case of a dominating antiferromagnetic RKKY interaction, the authors of Refs.\ \cite{PaulIngersent1996,Ingersent2005} still differentiate between two frustrated scenarios, which they called 'the 'frustrated Kondo regime'' and ''isospin Kondo-regime'', both characterized by NFL FPs 
(for further explanation see Refs.\ \cite{PaulIngersent1996} and \cite{Ingersent2005}).
Whereas the isospin Kondo-regime was found to be unstable against weak PH asymmetry, the frustrated Kondo regime is robust against moderate PH asymmetry.

In order to study the model with a $C_3$ symmetry
within our effective low-energy mapping, we applied the NRG with a discretization of $\lambda=4$ and kept $N_s=3000$ states after each diagonalization in the NRG procedure. 
Note, that even if $N_s=3000$ states might be insufficient to calculate thermodynamic quantities within a three-channel NRG calculation, the authors of Ref.\ \cite{PaulIngersent1996} only kept $N_s=1200$ states in their NRG calculations, which was sufficient to reproduce a conformal field theoretical description \cite{Ingersent2005} of the model.

Due to the $C_3$ symmetry, two independent parameters determine the influence of the original host conduction band onto the dynamics of the correlated orbitals in the effective low-energy model, $\Gamma^\text{off}=\Gamma^\text{off}_{12}=\Gamma^\text{off}_{23}$ and $t^\text{eff}=t^\text{eff}_{12}=t^\text{eff}_{23}$. In this section we focus on the parameter regime $-0.5<\Gamma^\text{off}/\Gamma_0<1$, such that 
rank$(\mat{\Gamma}_\sigma)=3$ is satisfied, and we are always studying a MIAM of the first kind. 

The case of dominating ferromagnetic RKKY interaction, or $\Delta R\to\infty$ respectively, corresponds to small $t^\text{eff}\to 0$ and results in the fully Kondo-screened FL fixed point, independently of the choice of $\Gamma^\text{off}$.

For dominating AF RKKY interactions, corresponding to an appropriate value of $t^\text{eff}$, we can still differentiate between two situations:
Applying $C_3$ point group properties,
two of the three eigenvalues of $\mat{\Gamma}_\sigma$ are identical. 
These are associated with the two helical states which are complex conjugated to each other.
 For $\Gamma^\text{off}/\Gamma_0<0$, these two eigenvalues define the larger couplings in the effective low-energy model, whereas for $\Gamma^\text{off}/\Gamma_0>0$, the single eigenvalue dominates.
$\Gamma^\text{off}/\Gamma_0=0$ implies an independent conduction band for each impurity in real space and, consequently, to three identical eigenvalues $\Gamma_n=\Gamma_0$.

In order to artificially restore PH symmetry, which is necessary to reproduce both NFL fixed points found in Refs.\ \cite{PaulIngersent1996,Ingersent2005}, we replace the effective tunneling elements $t^\text{eff}$ by an effective Heisenberg exchange interaction $J=4(t^\text{eff})^2/U$ between the correlated orbitals
\begin{align}
\sum_{\sigma}\sum_{ij, i\not=j}t^\text{eff}f^\dagger_{i\sigma}f_{j\sigma}\to \sum_{ij, i\not=j}J\vec{S}_i\vec{S}_j.
\label{eq:restorePH}
\end{align}
This incorporates the correct AF part of the RKKY interaction but removes the PH-symmetry breaking term by hand.
For this PH symmetric setting, and $J/T_K\gg1$, we can identify the isospin Kondo-regime for $\Gamma^\text{off}/\Gamma_0\leq0$ and the frustrated Kondo FP for $\Gamma^\text{off}/\Gamma_0>0$. Whereas the isospin Kondo FP  is unstable against small deviations from the PH symmetric point, $\e^f\to\e^f\pm\delta\e$, the frustrated Kondo FP remains stable in accordance with Refs.\ \cite{PaulIngersent1996,Ingersent2005}.
However, our NRG calculations reveal, that even the frustrated Kondo FP is unstable when a sufficient tunneling element $t>t_c$ between the correlated orbitals is added.

This finding has profound consequences for the NFL fixed point, since such a tunneling matrix
element is always dynamically generated in the full model, if the impurities are coupled to the same conduction band \cite{Eickhoff2018}.
The strength of RKKY interaction and potential scattering in the conduction band channels can not be treated 
as independent parameters, as done in Refs.\ \cite{PaulIngersent1996,Ingersent2005}.
In our effective low-energy model the AF part of the RKKY interaction, as well as the potential scattering, is generated by the same parameter $t^\text{eff}$.

\begin{figure}[t]
\begin{center}

\includegraphics[width=0.45\textwidth]{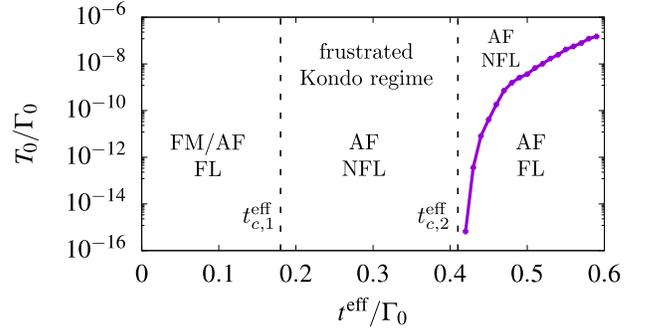}

\caption{Phase diagram for the low-energy model with $C_3$ symmetry as function of the tunneling element $t^\text{eff}$ between the correlated orbitals and $\Gamma^\text{off}/\Gamma_0=0.2$, $U/\Gamma_0=20$, $\e^f/\Gamma_0=-10$, $D/\Gamma_0=10$. Small values of $t^\text{eff}$ result in a FM or weak AF exchange interaction and a FL. Intermediate $t_{c,1}^\text{eff}<t^\text{eff}<t_{c,2}^\text{eff}$ lead to an AF exchange and the frustrated Kondo regime. For large $t^\text{eff}>t^\text{eff}_{c,2}$ the PH asymmetry destroys the NFL fixed point on a crossover scale (purple linepoints) which is exponentially small for $(t^\text{eff}-t_{c,2}^\text{eff})\to 0^+$.
 }
\label{fig:C3_T0}
\end{center}
\end{figure}

In order to study the phase diagram of our mapped 
$C_3$ symmetric MIAM, we recall that due to the symmetry restrictions we have
only two independent parameters $\Gamma^\text{off}$ and $t^\text{eff}$, stemming 
from the real and imaginary part of the self-energy matrix $\mat{\Delta}_\sigma(z)$.
We keep the ratios $U/\Gamma_0$ and $\Gamma^\text{off}/\Gamma_0>0$ fixed in order
to access the frustrated Kondo-regime of the model. 
We summarized the low-energy fixed points found in the NRG calculation for
$U/\Gamma_0=20$ and $\Gamma^\text{off}/\Gamma_0=0.2$ 
as a function of the remaining free parameter $t^\text{eff}$
in Fig.\ \ref{fig:C3_T0}.

\begin{table}[b]
	\centering
		\begin{tabular}{c|c|c|c}
			  & $\Gamma^\text{off}/\Gamma_0=0.2$ & $\Gamma^\text{off}/\Gamma_0=0.4$ & $\Gamma^\text{off}/\Gamma_0=0.6$ \\ \hline
 	\multirow{2}{*}{$U/\Gamma_0=20$} & $t^\text{eff}_{c,1}/\Gamma_0\approx 0.18$ & $t^\text{eff}_{c,1}/\Gamma_0\approx 0.29$ & $t^\text{eff}_{c,1}/\Gamma_0\approx 0.61$ \\
	& $t^\text{eff}_{c,2}/\Gamma_0\approx 0.41$ & $t^\text{eff}_{c,2}/\Gamma_0\approx 0.96$ & $t^\text{eff}_{c,2}/\Gamma_0\approx 1.94$  \\ \hline
	$U/\Gamma_0=10$ & - & - & -\\
		\end{tabular}
	\caption{Critical strengths $t_{c,1}^\text{eff}$ and $t_{c,2}^\text{eff}$ of the low-energy model with $C_3$ symmetry and PH symmetric impurities, for three different values of $\Gamma^\text{off}/\Gamma_0$ and two different Coulomb interactions $U/\Gamma_0$. For $U/\Gamma_0=10$ the NFL fixed point is absent for all regarded $\Gamma^\text{off}/\Gamma_0$.}
	\label{tab:tc}
\end{table}

The impurity contribution to the entropy, $S_{\rm imp}$ \cite{BullaCostiPruschke2008}, is a  measure for
the fixed point properties and is related to the ground state degeneracy $g$ of simple
low-energy fixed points, $S_{\rm imp} = k_B \ln(g)$. In the local moment regime, $g$ is given by an integer value
$g=2S+1$, where $S$ is the effective spin of the local moment. At NFL FPs, however, $g$ can also acquire
an irrational number  parametrizing the unconventional low-energy
excitation spectrum at the FP \cite{Ingersent2005}.

Since a finite $\Gamma^\text{off}$ is responsible for
a FM RKKY interaction,
we find the fully Kondo-screened FL FP for small $t^\text{eff}$.
The spin-spin correlation function starts from  ferromagnetically  aligned spins for $t^\text{eff}=0$
and is reduced with increasing $t^\text{eff}$. Shortly before the critical values $t^\text{eff}_{c,1}$
the correlation function changes its sign to weakly AF-coupled local moments.
At a critical $t^\text{eff}_{c,1}/\Gamma_0\approx0.18$, the AF part of the RKKY interaction dominates, leading to the frustrated Kondo regime with irrational degeneracy $g$ of the ground state \cite{AffleckLudwig1991}, 
$g=[1/2(5+\sqrt{5})]^{0.5}\approx  1.9$, as reported in \cite{Ingersent2005}.

While further increasing $t^\text{eff}$, we observe a second critical value $t^\text{eff}_{c,2}/\Gamma_0\approx0.41$ at which the NFL fixed point 
becomes unstable: For $t^\text{eff}>t^\text{eff}_{c,2}$,
the system flows to the unstable frustrated Kondo fixed point at intermediate temperatures, but crosses over to a stable FL below the low-energy scale $T_0$ (purple line points in Fig.\ \ref{fig:C3_T0}), which is defined at the point where the entropy reaches the midpoint between both fixed points, $S_{\rm imp}(T_0)=1/2(S_{\rm imp}^\text{FL}+S_{\rm imp}^{\text{NFL}})$.
This new crossover scale is exponentially suppressed when
approaching the QCP for $(t^\text{eff}-t^\text{eff}_{c,2})\to 0^+$.

In order to find a finite interval of $t^\text{eff}$ in which the NFL fixed point is stable,
the total RKKY coupling $K_\text{RKKY}$, comprising 
the FM and AF contributions $K_\text{RKKY}=K_\text{RKKY}^\text{FM}+K_\text{RKKY}^\text{AF}$
needs to fulfill two conditions:
(i) the AF part of the RKKY interaction has to dominate over the FM one, $K_\text{RKKY}^\text{AF}>K_\text{RKKY}^\text{FM}$, and, (ii) $K_\text{RKKY}$ needs to be larger than the single-ion Kondo temperature but must not exceed an upper critical value
\begin{align}
T_\text{K}<K_\text{RKKY}<K_\text{RKKY}^c\,,
\label{eq:Kc}
\end{align}
since an increasing $K_\text{RKKY}^\text{AF}$ is associated with increasing potential scattering destroying the NFL fixed point.
As a result of the upper bound in Eq.\ \eqref{eq:Kc}, the NFL fixed point is absent once $T_K>K_\text{RKKY}^c$.

\begin{figure}[t]
\begin{center}

\includegraphics[width=0.45\textwidth]{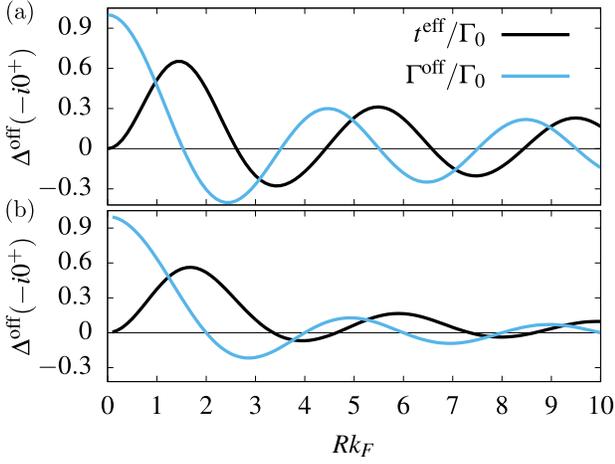}

\caption{Effective tunneling element $t^\text{eff}=\rm{Re}\Delta^\text{off}(-i0^+)$ (black) and $\Gamma^\text{off}=\rm{Im}\Delta^\text{off}(-i0^+)$ (light blue) as function of the dimensionless distance $Rk_F$ for an isotropic linear dispersion $\e_{\k}$ in (a) 2d and (b) 3d. 
 }
\label{fig:C3_RKKY}
\end{center}
\end{figure}

We repeated the NRG calculations for two other ratios of $\Gamma^\text{off}/\Gamma_0$
as well and find the same sequence of low-energy fixed points.
We summarize the two critical  $t^\text{eff}_{c,1/2}$ for three different values of $\Gamma^\text{off}/\Gamma_0$ and two different Coulomb interactions $U/\Gamma_0$, for PH-symmetric impurities, i.e., $\e^f=-U/2$,
in table\ \ref{tab:tc}.
The larger the value of $\Gamma^\text{off}/\Gamma_0$, the larger is the interval 
$(t^\text{eff}_{c,2}-t^\text{eff}_{c,1})$ of the NFL regime as is apparent for $U/\Gamma_0=20$. 
Since a large $\Gamma^\text{off}$ increases the FM RKKY interaction, 
a larger $t^\text{eff}>t^\text{eff}_{c,1}$ is necessary
in order to reach the stable frustrated Kondo regime.

In the case of $U/\Gamma_0=10$, the single-ion Kondo temperature
is too large to fulfill Eq.\ \eqref{eq:Kc}, and the NFL fixed point is absent for all three values of $\Gamma^\text{off}/\Gamma_0$ listed in table\ \ref{tab:tc}.
Consequently, a small Kondo temperature in combination with large $\Gamma^\text{off}$ \textit{and} $t^\text{eff}$ stabilizes the NFL fixed point over a large region in the parameter space.

However, $\Gamma^\text{off}$ and $t^\text{eff}$ cannot be chosen independently
in real materials,
since they  result from the imaginary and real part of the same 
complex hybridization matrix $\mat{\Delta}_\sigma(-i0^+)$.
In Fig.\ \ref{fig:C3_RKKY}, we plotted $\Gamma^\text{off}$ (light blue) and $t^\text{eff}$ (black) in (a) 2d and in (b) 3d, as a function of the dimensionless distance $R k_F$, using an isotropic linear dispersion $\e_{\k}$ for the
host conduction band.
Due to the typical phase shift between $\Gamma^\text{off}$ and $t^\text{eff}$, large values of $t^\text{eff}$ in general correspond to small values of $\Gamma^\text{off}$, destabilizing the NFL fixed point.

In conclusion, our low-energy mapping demonstrates that it indeed is possible to realize the NFL fixed point of the frustrated Kondo regime found in Refs.\ \cite{PaulIngersent1996,Ingersent2005}. Nevertheless, aside from the 
$C_3$-symmetric setup,  this requires a substantial small single-ion Kondo temperature and a small impurity separation $R k_F\approx 1$ in order to achieve an appropriate combination of $\Gamma^\text{off}/\Gamma_0>0$ and $t^\text{eff}_{c,1}<t^\text{eff}<t^\text{eff}_{c,2}$.

\subsection{Multi-impurity problems connected to a 1d host}
\label{sec:MIAM_1d}

In this section we present the results for several MIAMs of the second kind by applying the NRG to the effective low-energy models.
We focus on  the low-energy fixed points, the crossover temperatures and spin-spin correlation function
for  $N_f=3,4$ and $N_f=5$ as a function of the band filling, and identify  a series of 
Kosterlitz-Thouless type  QCPs  that are associated with 
a change of the fixed point degeneracy.  Since the decay of the effective hopping matrix element with the impurity-impurity distance
is the slowest in 1d, we used a 1d host to maximize the magnetic frustration induced by hopping matrix
${\rm Re}\mat{\Delta}_\sigma$. 
Note that one of the main results, namely the existence QCPs in the phase diagram, which emerge due to FM correlations between the local moments, is \textit{not} restricted to 1d systems.
These QCPs originate from the fact, that we are studying MIAMs of the second kind with 
${\rm rank}(\mat{\Gamma}_\sigma)\leq 2 < N_f$, which likewise exist in higher spatial dimensions as discussed in Sec.\ \ref{sec:Rank2d3d}.
We also use a simple nearest-neighbor tight binding model with a dispersion $\e_{k\sigma} = -2t \cos(k a) +\e_c$. 
In  the case of an odd number $N_f$ we prevent a spin singlet formation already within the multi-impurity cluster.

The $N_f$ impurities  form a short finite-size chain in 1d. The orbital energies and the Coulomb repulsion $U$ 
are chosen such that the local orbitals  are approximately singly occupied at low
temperature, and the induced local moment is subject to the remaining  
interactions. 

By shifting the band center $t_{ii} = \e_c$ from $\e_c=0$ to  $\e_c\to D=2t$, we reduce
the electron filling in the
conduction band of the host, reduce the Fermi wave vector $k_f$, and  hence altering the matrix elements  
$\Delta_{lm,\sigma}(-i0^+)$
such that the spin-spin correlation functions are associated with a longer wave length.
While $\e_c=0$ refers to a half-filled particle-hole symmetric conduction band, we approach the
one conduction electron limit
for $|\e_c|/D\to 1$ that has been considered by Sigrist et al.\ \cite{KLM-Sigrist91}.
The analysis of  this 1d finite size impurity chain problem shows that ${\rm rank}(\mat{\Gamma}_\sigma)=2$,
with the exception of $|\e_c| = D$ where the rank changes to one.

Three points are worth  noting.
Such problems (i) can be treated by a two-band NRG approach as a consequence of ${\rm rank}(\mat{\Gamma}_\sigma)\le 2$  and are limited only
by the maximally manageable dimension of the local cluster Hilbert space. (ii) The cluster contribution \cite{Wilson75}  to the low-energy fixed point entropy, $S_{\rm imp}$, 
 is given by $S_{\rm imp}=k_B \ln(N_f)$ in the limit $|\e_c| \to  D$ in accordance with Ref.\ \cite{KLM-Sigrist91}. 
For $N_f>2$, we are (iii) always investigating a MIAM problem of the second kind 
depicted as regime III in Fig.\ \ref{sec:categories-MIAMs}.

In this section we start with dense impurity arrays where the correlated orbitals are placed next to each other, such that the antiferromagnetic RKKY interaction dominates for a roughly half-filled conduction band. We study the effect of frustrated RKKY interaction at intermediate band-fillings and QCPs due to FM correlations occuring near the band-edges for PH symmetric impurities and at intermediate fillings for large PH asymmetry. Following this, we present NRG results for dilute impurity configurations with dominating ferromagnetic RKKY interactions for a half-filled conduction band, which is more suited for ferromagnetic HF materials.
These dilute multi-impurity models exhibit FM correlations and the associated QCPs in a realistic region of the parameter space which might be connected to those found in the quasi-1d ferromagnetic HF materials \cite{FMQCP_Kotegawa2019,FMQCP_Shen2020,FMQCP_Okano2015}. 

Unless otherwise stated, we set the NRG
discretization parameter $\lambda=3$ and kept $N_S=5000$ states after each iteration.

\subsubsection{Kosterlitz-Thouless type  Quantum Phase transitions and the Phase diagram in dense impurity arrays}
\label{sec:KT-QPT}

\begin{figure}[tb]
\begin{center}
\includegraphics[width=0.45\textwidth]{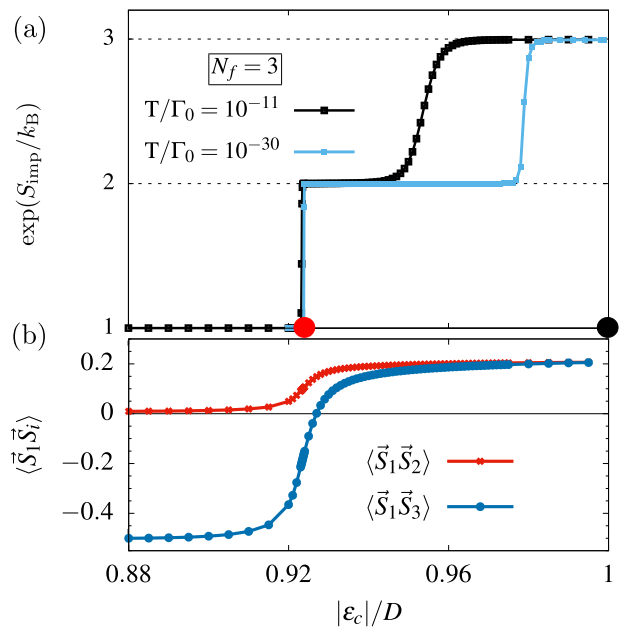}

\caption{(a) Impurity contribution to the entropy for $T\to 0$ 
as function of the band center for three particle-hole symmetric impurities
with $U/\Gamma_0=10, \e^f_i=-U/2$ and $D/\Gamma_0=10.$
(b)  The two  spin-spin correlation functions for the same parameter. The center impurity is labeled $i=2$, the two outer correlated
sites $i=1,3$. }
\label{fig:2-miam-Nf-3}
\end{center}
\end{figure}

Above, we introduced the parametrization of 
the impurity contribution to the entropy, $S_{\rm imp}$,
in terms of a  ground state degeneracy $g$: $S_{\rm imp} = k_B \ln(g)$. 
This effective degeneracy $g= \exp(S_{\rm imp}/k_B)$
is plotted as function of $|\e_c|$ in Fig.\ \ref{fig:2-miam-Nf-3}(a) for $N_f=3$
and a particle-hole symmetric Hamiltonian $H_{\rm corr}$. It 
shows the typical behavior of the MIAM under investigation here. Starting from a singlet
ground state, $g=1$, the degeneracy increases  in integer steps to the maximum $g=N_f$, which is always reached for 
$|\e_c|/D\to 1$. The  point of increase defines a quantum critical point (QCP)
which is  of Kosterlitz-Thouless (KT) type as shown below.

In Fig.\ \ref{fig:2-miam-Nf-3},  we concentrate on parameters
close to the changes of entropy. In order to understand the physics of the two different quantum phases in the depicted region of $\e_c$, the two different spin-correlation functions
are plotted   in Fig.\ \ref{fig:2-miam-Nf-3}(b).

In region I, $S_{\rm imp}(T\to 0)$ vanishes, and the correlation functions
of the two outer impurity spins are anti-parallel while the central spin moment is screened by the Kondo effect. This setup is equivalent to the triangular spin cluster investigated in Refs.\ \cite{Schnack2019,MitchellLogan2009,MitchellLogan2013}.  The correlation function between the two outer spins with the central spin almost vanishes.  The spin singlet formation involves two components: the AF interaction mediated by the dominating next-nearest hopping between two outer spins, and the Kondo effect in the even sector that removes the local moment of the central spin.

The QCP occurs at the red point added to the horizontal axis in Fig.\ \ref{fig:2-miam-Nf-3}(a): At this point $S_{\rm imp}$ jumps to $k_B\ln(2)$ indicating a doubly degenerate ground state in region II. 
The spin-spin correlation function, however, shows a smooth crossover at this point 
indicating the physics of the new ground state.
The nearest neighbor spin-spin correlation function depicted in red in  Fig.\ \ref{fig:2-miam-Nf-3}(b) rises from an anti-alignment 
to a finite ferromagnetic correlation. This observation is consistent  with the notion that a reduction of $k_f$ eventually changes the nearest neighbor RKKY interaction from AF to FM. 
The next-nearest neighbor spin-spin correlation function changes sign to ferromagnetic correlations of the same magnitude: 
All three local moments  of the correlated cluster
are  ferromagnetically coupled in this parameter regime and form a large $I=3/2$ 
total spin at the unstable local moment fixed point.
Since we only have two  effective conduction electron band channels 
due to ${\rm rank}(\mat{\Gamma}_\sigma)=2$, we are in an underscreened Kondo regime. 
We found a two-stage Kondo effect quenching the $I=3/2$ local spin down to $I=1/2$ via the two conduction electron channels coupling with different coupling constants. This 
is consistent with the residual entropy of a remaining decoupled doublet in the stable low-energy fixed point.

Note that the FM RKKY coupling is essential to stabilize the LM fixed point.
In Refs.\ \cite{MitchellLogan2009,MitchellLogan2013} the authors studied a quantum dot trimer consisting of a specific symmetry in real space, such that only the subspace with even parity is directly coupled to a conduction band screening channel. In this special case it is possible to also use 
the values of single-particle hopping terms between the orbitals and the hopping-induced AF
coupling between the local moments
to stabilize a local moment with odd parity. However, if the local moments are coupled via the RKKY interaction, i.e., 
all orbitals hybridize with one single conduction band in real space, each irreducible subspace contains at least one screening channel in general and, consequently, it is impossible to stabilize the LM fixed point for AF RKKY couplings.

The rank of the $\Gamma_\sigma$-matrix changes to ${\rm rank}(\mat{\Gamma}_\sigma)=1$
at the point $|\e_c|/D=1$: the total 
$I=3/2$ spin formed by the FM RKKY interaction defining the unstable local moment fixed point
can only be screened by the single remaining conduction electron channel leading to the cluster entropy of $S_{\rm imp}/k_B =\ln(3)$. 

The  black entropy curve in  Fig.\ \ref{fig:2-miam-Nf-3}(a)  suggests that this is a smooth transition at finite temperature, 
governed by a  crossover value $|\e_c |/D<1$. 
This data, however, was obtained for a fixed number of NRG iterations that
corresponds to a fixed temperature of $T/\Gamma_0 = 10^{-10}$. We increased the number of NRG iterations and added the results as the light-blue curve representing a temperature that is 20 orders of magnitude lower, i.\ e., 
$T/\Gamma_0 = 10^{-30}$. The crossover region 
is clearly pushed  closer to $|\e_c |/D=1$.

The finding in region II of  Fig.\ \ref{fig:2-miam-Nf-3}(a) can be understood in terms of 
a band filling dependent low-energy scale $T_{L}(\e_c)$ that governs the crossover 
from the unstable local moment fixed point with $I=3/2$
to the stable low-energy fixed point which is characterized by $S_{\rm imp} =k_B\ln(2)$.
This low-energy scale can be associated with the Kondo-screening due to the smaller coupling $\Gamma_n$ that vanishes for $|\e_c|\to D$.
To access the entropy of the low-energy fixed point requires that $T\ll T_{L}(\e_c)$. 
From our NRG data we can conclude that 
\begin{eqnarray}
\lim_{|\e_c|\to D} T_{L}(\e_c) &=&0
\end{eqnarray}
and, therefore, the critical point (i) is located at $|\e_c |/D=1$ and (ii) 
is  of Kosterlitz-Thouless type
since it vanishes as $T_{L}(\e_c) \propto \exp(A/\sqrt{D-|\e_c|})$ with some fitting parameter $A$.

\begin{figure}[t]
\begin{center}
\includegraphics[width=0.46\textwidth]{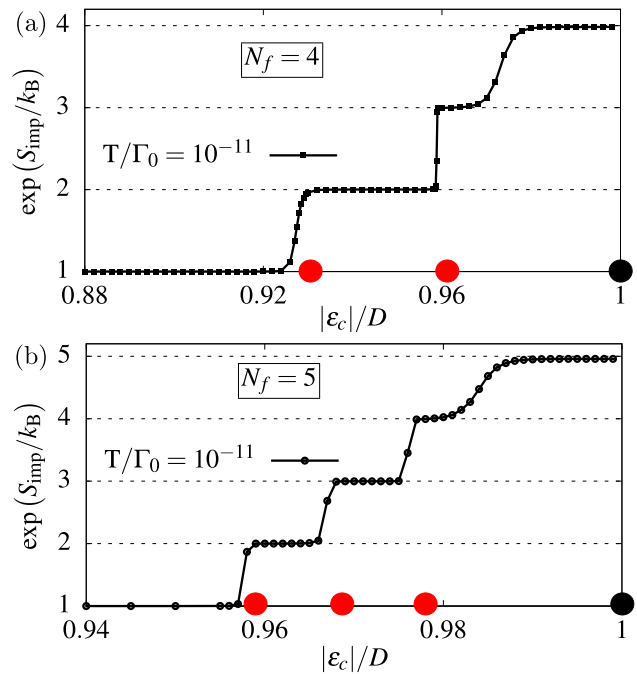}

\caption{$S_{\rm imp}$ as function of $|\e_c|/D$ for (a) $N_f=4$ and (b) $N_f=5$ 
and a fixed temperature $T/\Gamma_0=10^{-11}$ (corresponding to a fixed number of NRG iterations $N=50$.) for particle-hole symmetric impurities with $U/\Gamma_0=10$.
 For $N_f=4$ we find three regions and three KT quantum phase transition points, for $N_f=5$ four
regions separated by a KT-type phase transition are identified. 
}

\label{fig:mian-nf-45-Simp}
\end{center}
\end{figure}

The investigation of the low-energy fixed points of the MIAM with $N_f=4$ and $N_f=5$ reveals
a similar picture as shown in Fig.\ \ref{fig:mian-nf-45-Simp}: for a half-filled conduction band
we always find a vanishing $S_{\rm imp}$ for the stable low-temperature fixed point 
that can be interpreted as a spin-singlet ground state formation.
In the limit $|\e_c|\to D$, we reproduce the prediction of Sigrist et al.~\cite{KLM-Sigrist91,Nagaoka1966} even in our finite size system: 
an unstable local moment fixed point with $I=N_f/2$ is formed via FM RKKY interactions,
and is screened by the Kondo effect to the stable fixed point spin $I=(N_f-1)/2$,
corresponding to  $S_{\rm imp}= k_b\ln(N_f)$, since   ${\rm rank}(\mat{\Gamma}_\sigma)=1$ 
at this point independently of $N_f$. The ground state degeneracy $g$ rises in integer steps from 
$g=1$ to $g=N_f$,  and the corresponding QPTs are of KT type.

In each of the regions one additional moment is aligned ferromagnetically with the others: The self-screening of the local impurity spins 
via the RKKY mediated interaction becomes less and less effective since this interaction becomes predominantly ferromagnetic for low band fillings.
The hopping matrix elements ${\rm Re}\Delta_{lm}(\e_c)$ reach their maximum at $|\e_c|=D$ due to the maximal particle-hole asymmetry of the conduction band. At the same time, the density of states diverges at the band edge in 1d, such that the hybridization to the conduction bands start to dominate
resulting in a FM alignment of the spins \cite{Jones_et_al_1987,Eickhoff2018}.

Depleting the conduction band more and more, leaves asymptotically a Hubbard type
model \cite{PruschkeHM1990} where close to half-filling the FM-aligned
local moments could be interpreted as a precursor of the Nagaoka mechanism \cite{Nagaoka1966} to ferromagnetism in the Hubbard model. Energy is gained by allowing the impurity electrons to delocalize 
in a uniform polarized background.

\begin{figure}[t]
\begin{center}
\includegraphics[width=0.45\textwidth]{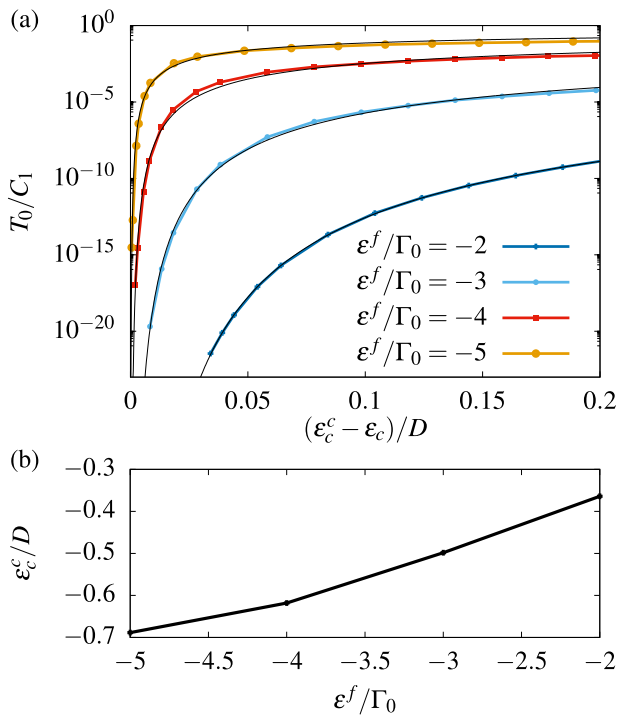}
\caption{(a) Low-temperature scale  $T_0$
for particle-hole asymmetric three-impurity models
with $U/\Gamma_0=50,D/\Gamma_0=10$, and four different values of $\e^f$.
(b) The critical value $\e_c^c$ as a function of $\e^f$.
 }
\label{fig-1-9}
\end{center}
\end{figure}

We still need to prove the claim that the transitions between the different regions in
 Fig\ \ref{fig:2-miam-Nf-3}(a) and Fig.\ \ref{fig:mian-nf-45-Simp}, indicated by the red dots on the
horizontal axis, are indeed QCPs of the KT type.
We exemplify this point by investigating the low-temperature scale $T_0(\e_c)$ as
a function of $\e_c$ in the three impurity problem, $N_f=3$, for four different values of $\e^f$ and a fixed $U/\Gamma_0=50$. We  selected these parameters to demonstrate  that the transition type is unrelated to the particle-hole symmetry
and that the critical value of $\e^c_c$ depends on the particle-hole asymmetry of the 
correlated cluster. 
In Ce, for instance, the $4f$-shell occupation fluctuates between zero and one, so that  the limit $U\to\infty$  was successfully employed \cite{Grewe83,Kuramoto83,Grewe87} to understand
the basic properties of such systems. 

The low-temperature scale $T_0(\e_c)$ is defined via the crossover from
the last unstable local moment fixed point to the stable low-temperature fixed point
when $S_{\rm imp}(T)$ reaches the mid point between both entropies: 
\begin{eqnarray}
S_{\rm imp}(T_0) = \frac{1}{2}(S^{\rm LM}_{\rm imp}+S^{\rm LT}_{\rm imp}(0)).
\end{eqnarray}
We fitted the $T_0(\e_c)$ to the exponential form 
\begin{eqnarray}
\label{eq:KT-fit}
T_0(\e_c)= C_1 e^{\frac{C_2}{\sqrt{\e_c^c-\e_c)}}}
\end{eqnarray}
close to the transition point
and extracted the transition point $\e_c^c$ as well as the parameters $C_1$ and $C_2$.
We plotted the NRG data $T_0(\e_c)/C_1$ as a function of $(\e_c^c-\e_c)/D$
in Fig.\ \ref{fig-1-9}(a). In addition we added the fitting  curve defined in Eq.\ \eqref{eq:KT-fit}
as black solid lines for the four cases demonstrating
an excellent agreement with the NRG data with this analytical form.
This can be done in the vicinity of all QCPs for different cluster sizes $N_f$.
Therefore, all QCPs at the critical values indicated by red or black dots in
Figs\ \ref{fig:2-miam-Nf-3}(a) and Fig.\ \ref{fig:mian-nf-45-Simp}
are of the KT universality class.

We added the particle-hole asymmetry dependency of $\e_c^c$ as Fig.\ \ref{fig-1-9}(b).
Upon increasing $\e^f$ from $\e^f/\Gamma_0=-5$ to $\e^f/\Gamma_0=-2$, $\e_c^c$ is significantly reduced.

\subsubsection{The role of magnetic frustration in the $S_{\rm imp}=0$  phase}
\label{sec:kinks-magnetic-frustration}

After establishing a step wise increase of the fixed point degeneracy $g$
in the limit $|\e_c|/D\to 1$ (vanishing electron or hole conduction band filling,)
we focus  on the largest  interval in the $\e_c$ parameter space in this section.
This regime is determined by $S_{\rm imp}=0$, indicating a singlet ground state
of the low-energy fixed point.

The notion of a competition between an AF RKKY screening
and the Kondo screening by the two conduction channels  was established
in the two-impurity model \cite{Jones_et_al_1987}. In this MIAM of first kind, there are always enough
conduction electron channels available to Kondo-screen all the local moments.
For $N_f>2$ and  ${\rm rank}(\mat{\Gamma}_\sigma)<N_f$, the topology of the model becomes different and
such a scenario is not  applicable any more. The concept of competing Kondo effect and RKKY interaction, both generated by
the host conduction electrons, become less meaningful due to the lack of available screening channels.
For periodic systems, Eq.\ \eqref{eq:delta-mat-def} suggests that the ${\rm rank}(\mat{\Gamma}_\sigma)$ is of the order of the $\k$-points on the Fermi surface and, therefore, is much smaller then $N_f$.

\begin{figure}[t]
\begin{center}
\includegraphics[width=0.45\textwidth]{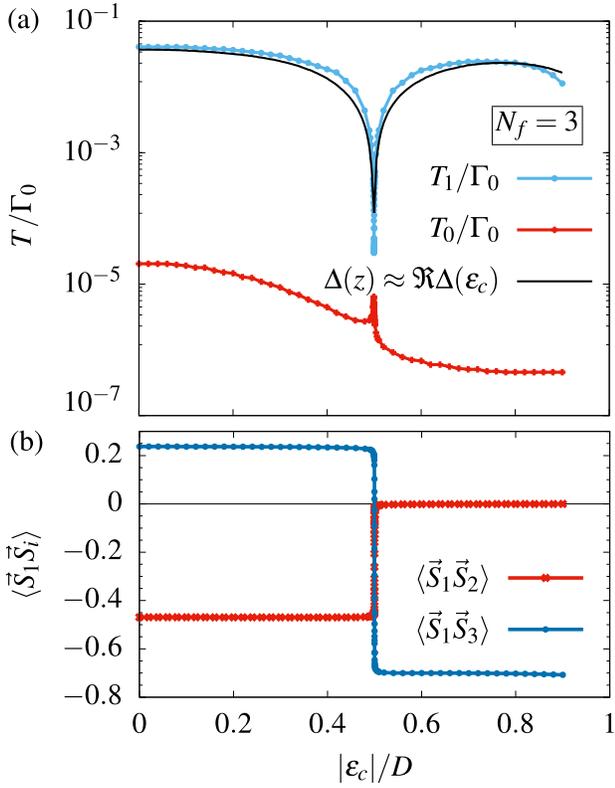}
\caption{(a) The two low-temperature scales $T_0$ and $T_1$ vs $\e_c$ in the three-impurity model.
$T_1$ (blue) is defined as crossover temperature approaching the unstable LM fixed point
with $I=1/2$ and $T_0$ (red) denotes the  crossover temperature from the unstable $S=1/2$ LM fixed point to the singlet fixed point. The crossover temperature for the multi-impurity system without coupling to the conduction band channels is added as a black line.
(b) Low-energy fixed point spin-spin correlation functions $\expect{\vec{S}_1\vec{S}_2}$
and  $\expect{\vec{S}_1\vec{S}_3}$ vs $\e_c$.
Parameters: particle-hole symmetric impurities with $U/\Gamma_0=30,\e^f_l=-U/2$, $D/\Gamma_0=10$.}
\label{fig6-Simp0-T0-spin-spin}
\end{center}
\end{figure}

We begin with the three impurity model and calculate the two low-energy 
crossover temperatures to the $S_{\rm imp}=0$ stable fixed point:
$T_1(\e_c)$ denotes the crossover temperature to the last unstable $S=1/2$ LM fixed point,
and $T_0(\e_0)$ characterizes the approach to the stable low-energy fixed point. The temperature $T_0$ replaces $T_K$ in the single-impurity model since it is associated with the Kondo effect.
The results are plotted in Fig.\ \ref{fig6-Simp0-T0-spin-spin}(a). We note that both crossover temperatures change continuously with $\e_c$ 
but develop a cusp at $|\e_c|/D=0.5$. $T_0$ remains finite over the whole parameter regime until the end at about $|\e_c|/D\approx 0.945$ where the KT-type QPT to a stable fixed point with $S_{\rm imp}/k_B=\ln(2)$ occurs that was discussed in the previous section. 
The overall decrease in $T_0(\e_c)$  is related to the dependency of the effective change of fluctuation strengths $\Gamma_{n\sigma}$ 
as function of $\e_c$: the coupling to the relevant orbital involved in the Kondo screening decreases
with increasing $|\e_c|$.

The spin-spin correlation functions calculated at the low-energy fixed point 
reveals that two different regions emerge as shown  in Fig.\ \ref{fig6-Simp0-T0-spin-spin}(b).
For small values $|\e_c|/D$, the nearest neighbor spins are AF aligned while the two outer impurity spin are FM correlated.  The spin-spin correlation functions indicate that the three local spins always
add up to a $S=1/2$ ground state at intermediate temperature due to the RKKY interaction. The approach to the ground state configuration
of the impurity cluster occurs on the temperature scale $T_1$. One of the two conduction electron channels is sufficient to Kondo screen the remaining $S=1/2$ distributed over all impurities. That occurs on the crossover temperature scale $T_0$ where the impurity entropy is removed. For $0.5<|\e_c|/D$, the two local spins connected to the same sublattice are AF aligned while the center spin is screened 
by the Kondo effect.

\begin{figure}[t]
\begin{center}
\includegraphics[width=0.45\textwidth]{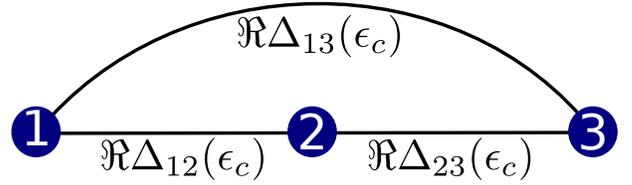}

\caption{Schematic diagram of the tight binding hopping parameters between the tree
impurities}
\label{fig:threeimp-hopping}
\end{center}
\end{figure}

In order to shed some light onto the nature of the observed cusp, $|\e_c|/D\approx 0.5 $,
we solved the  three impurity cluster 
model for $\mat{\Gamma}_\sigma=0$, i.\ e.\,by decoupling the impurities from
the two effective conduction bands, using exact diagonalization.
We calculated the crossover temperature to the low-energy ground state from the temperature dependence of the entropy $S_{\rm imp}(T)$.
This crossover scale was added as a continuous black line to Fig.\ \ref{fig6-Simp0-T0-spin-spin}(a) and traces
the NRG temperature $T_1$ very well.  
At the cusp, however, this crossover temperature vanishes in the decoupled cluster model: a level crossing between two different
twofold degenerate cluster ground states occurs which can be characterized by different spin-spin correlation 
functions.   This defined a point of maximal magnetic frustration in the system.

The nature of those local ground states can be understood in terms of the effective hopping
parameters in the mapped model as  shown schematically in Fig.\ \ref{fig:threeimp-hopping}.
Note that we drop the spin index for clarity in the following
since all parameters are spin-independent in the absence
of spin-polarized conduction band.
By only considering $\real \Delta_{13}(\e_c)$, we obtain from the orbitals 1 and 3, two molecular orbitals that have even or odd  symmetry under a parity transformation that can be constructed. The even orbital couples to the orbital 2 via $\real \Delta_{12}(\e_c)$ and $\real \Delta_{23}(\e_c)$ and two new even orbitals are generated. Then we fill these orbitals with three electrons. For $|\e_c|/D< 0.5$, 
one of the even orbitals has the lowest energy and is filled with two electrons forming a singlet. The third electron occupies
the orbital with odd symmetry. Therefore, the local moment is located at the outer edge on the orbitals 1 and 3 in real space.
For  $|\e_c|/D>0.5$, $\real \Delta_{13}(\e_c)$ dominates and the odd symmetry orbital has the lowest single-particle energy, and is doubly occupied. This singlet formation results in a strong AF correlation function $\expect{\vec{S}_1\vec{S}_3}$ in this regime.
The local moment is located in the higher-lying even orbital, and therefore, more localized in the orbital 2.

At the degeneracy point, all  three hopping matrix elements in Fig \ref{fig:threeimp-hopping} become equal and generate a maximally magnetically frustrated system. 
A level crossing between these  two doublet ground states occurs 
forming  a four-fold degenerate cluster  ground state.
This emerging local picture also explains the observed change in the spin-spin correlation function shown  in Fig.\ \ref{fig6-Simp0-T0-spin-spin}(b) for the full problem. For $|\e_d|/D>0.5$, the  
nearest neighbor spin correlations are suppressed and the central spin is Kondo screened.

Although the real-space geometry of our cluster is a short three-site chain, the hopping matrix elements ${\rm Re}\Delta_{ij}$
generate a trimer with degenerate AF Heisenberg couplings as investigated in the literature \cite{PaulIngersent1996,Ingersent2005,KoenigColemanKomijani2020,Schnack2019}.
The different physics found here is related to the reduced number of screening channels:
While Refs.\ \cite{PaulIngersent1996,Ingersent2005,KoenigColemanKomijani2020} investigate a Kondo trimer model of first kind, we derived an example of
a trimer model of the second kind \cite{Schnack2019}.

Right at the degeneracy point, both doublets with different parity symmetry 
couple to one of the two effective conduction bands.
If we assume PH symmetry and identical couplings $\Gamma_n$, this would result in the isospin Kondo regime found in Refs. \cite{PaulIngersent1996,Ingersent2005} and already discussed in Sec.\ \ref{sec:C3_frustration}. However, the effective tunneling elements ${\rm Re}\Delta_{ij}$, which generate the AF RKKY interaction and, therefore, are necessary in order to obtain magnetic frustration, break the PH symmetry and cause a FL with spin-singlet ground state, since this is a relevant perturbation of the isospin Kondo FP \cite{PaulIngersent1996,Ingersent2005}.

Furthermore, the mapped MIAM  lacks the helicity symmetry of the $C_3$ group in general, and, consequently, the two couplings $\Gamma_n$ of the degenerate doublets are not identical.
In this case, the cluster ground state degeneracy is lifted at the degeneracy point by the asymmetric coupling to the two conduction band channels, such that even the coupling of only one of the conduction electron channels to one of the two degenerate doublets can quench the remaining cluster entropy.
The energy splitting of both doublets, however, is dynamically generated by the RG procedure and remains small. Therefore,
we  do not observe a two-stage screening process: only one low temperature scale $T_0$ emerges, even at the degeneracy point, defining the crossover from the LM fixed point with $S_{\rm imp}\approx k_B\ln(4)$ to the stable strong coupling fixed point with $S_{\rm imp}=0$.

At the degeneracy point, the definition of $T_1$ becomes obsolete since we observe a direct crossover from a four-fold degenerate unstable local moment fixed point to the singlet ground state. Therefore, we do not find another
QCP in the $S_{\rm imp}=0$ regime.

While the low-energy scale $T_0$ decreases with increasing $|\e_c|/D$ due to the change in the eigenvalues $\Gamma_n$ that couple to the impurity ground state multiplet, the enhancement of the magnetic fluctuations in the vicinity of the degeneracy point causes the cusp in $T_0$ signaling the adiabatic change of the spin correlations in the low-energy fixed point.
The crossover temperature $T_0(\e_c)$ and the location of the cusp is symmetric in $\pm\e_c$, for particle-hole symmetric impurities,
qualitatively the same phases are observed for particle-hole asymmetric impurities but with asymmetric curves, not shown here.

\begin{figure}[t]
\begin{center}
\includegraphics[width=0.54\textwidth]{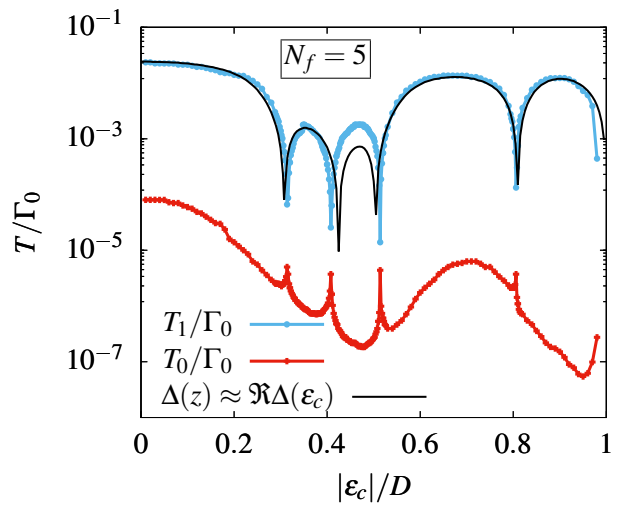}

\caption{The two low-temperature scales $T_0$ and $T_1$ vs $\e_c$ in the $N_f=5$ MIAM.
$T_1$ (blue) denotes crossover temperature approaching the lowest unstable LM fixed point
and $T_0$ (red)  crossover temperature from the unstable LM fixed point to the singlet fixed point. 
The crossover temperature for the multi-impurity system without coupling to the conduction band channels is added as a black line.
Parameters: as in Fig.\ \ref{fig6-Simp0-T0-spin-spin}
}

\label{fig8-Simp0-T0-Nf5}
\end{center}
\end{figure}

A similar picture emerges for $N_f=5$ in the $S_{\rm imp}=0$ regime, as shown in  Fig.\ \ref{fig8-Simp0-T0-Nf5}. For even $N_f$, the impurity cluster can form a singlet ground state
without coupling to the conduction bands: Hence we focus on odd $N_f$  where
the interplay between the RKKY interaction and the Kondo effect is relevant.
Clearly visible is the continuous change of $T_0(\e_c)$ but now showing  four cusps. 

Since the total spin is conserved in the model, the impurity cluster eigenstates
can be classified by the total angular momentum. The hopping matrix $\real{\Delta_{ij}(\e_c)}$ induces AF interactions between the spins, so that the cluster ground state is located in the subspace with the lowest angular moment. Adding  $N_f=5$ local moments $S=1/2$ yields five spin $J=1/2$ multiplets. Changing $\e_c$ is changing the energy spectrum of the five $J=1/2$ eigenstates, and, therefore, the nature of the cluster ground state within the subspace $J=1/2$.
The black solid line for $T_1$ added to  Fig.\ \ref{fig8-Simp0-T0-Nf5}
was calculated from the  impurity cluster spectrum including the Coulomb interaction but neglecting the coupling to conduction electrons channels in Eq.\ \eqref{eqn:MIAM_hyb}. 

The cluster calculation demonstrates that the cusps obtained from the full model 
are associated with a change
of the cluster ground state in the unstable intermediate fixed point. Note, however, that in the impurity cluster the scale $T_1$ vanishes at the four level crossing points  of the ground states, which is not 
visibly here since we did not zoom into the  degeneracy point with high resolution.

For the five different $J=1/2$ multiplets   four different cluster degeneracy points are found. At each degeneracy point, the two low-lying doublets couple to the two conduction electron bands. The screening of the impurity cluster entropy $S_{\rm imp}=k_B\ln(4)$ 
is characterized by the crossover temperature $T_0$.
Quantum fluctuations enhance $T_0$ at the degeneracy point and we also found, that a single spin degenerate band is sufficient to quench the local moments even at the degeneracy point
of the impurity cluster, where the local system is magnetically frustrated.

\subsubsection{Kosterlitz-Thouless type  Quantum Phase transitions and the Phase diagram in dilute impurity arrays}
\label{sec:KT-QPT_FM}

In dense impurity arrays, where the correlated orbitals are located  next to each other, the antiferromagnetic RKKY interaction typically dominates 
the interaction between the local moments of the correlated orbitals
for a roughly half-filled host conduction band.
In order to qualitatively simulate ferromagnetic HF materials \cite{FMKL_Krellner2007,FMKL_Tran2014,FMKL_Szlawska2018,FMQCP_Okano2015,FMQCP_Yang2020,FMQCP_Kirkpatrick2020,FMQCP_Kotegawa2019,FMQCP_Larrea2005,FMQCP_Khan2016,FMQCP_Shen2020,FMQCP_Steppke2013}, we 
study the dilute impurity configuration already discussed in Sec\ \ref{sec:dilute_model} and schematically depicted in Fig.~\ref{fig:2:depleted-PAM}. 
In this case, the
correlated orbitals only 
hybridize with the host Wannier orbitals of one of two sublattices in a bipartite lattice.
In Sec.~\ref{sec:dilute_model} we ascribed the ferromagnetic ground state at half-filling, $\e_c=0$,
obtained by
DMRG calculations \cite{Titvinidze2015,Potthoff2014,Potthoff2015,Potthoff2013},
to the absence of an effective tunneling between the correlated orbitals $\real\mat{\Delta}_\sigma(\e_c=0)=0$ and 
${\rm rank}(\mat{\Gamma}_\sigma)<N_f$. 

In the non-interacting limit ($U=0$),
the $d$-dimensional, depleted PAM can be exactly diagonalized, resulting in three bands. 
At particle-hole symmetry, corresponding to half-filling, one of these bands is totally flat, leading to a high degeneracy of the ground state, and possesses mainly $f$-character for small couplings $V/D$. As demonstrated by a first-order perturbation theory in $U$,
weak interactions within this flat band result in a fully polarized ground state of the model \cite{Potthoff2014}. For 1d and 2d, it was further shown that this polarized state persists to arbitrary strengths of the Coulomb interaction $U$ \cite{Titvinidze2015,Potthoff2014,Potthoff2015,Potthoff2013}.

From the perspective of our low-energy mapping, this result is just a consequence of the absence of the delocalizing, band generated effective hopping matrix elements
$\text{Re}\Delta_{ij,\sigma}(\e_c=0)=0$ 
in combination with an extreme reduction of available conduction band screening channels due to 
rank$(\mat{\Gamma}_\sigma)=1$.
However, for any $\e_c\not=0$, the flat band becomes dispersive, tantamount with the appearance of delocalizing tunneling elements $\text{Re}\Delta_{ij,\sigma}(\e_c\not=0)\not=0$ in the mapped Hamiltonian.
These delocalizing terms might lead to a localized/delocalized Mott-Hubbard insulator transition in the strongly interacting limit of the depleted PAM, 
with a possible connection 
to the emerging NFL behavior at the FM QCP in HF ferromagnets \cite{FMQCP_Kotegawa2019,FMQCP_Shen2020,FMQCP_Okano2015}.

\begin{figure}[t]
\begin{center}

\includegraphics[width=0.45\textwidth]{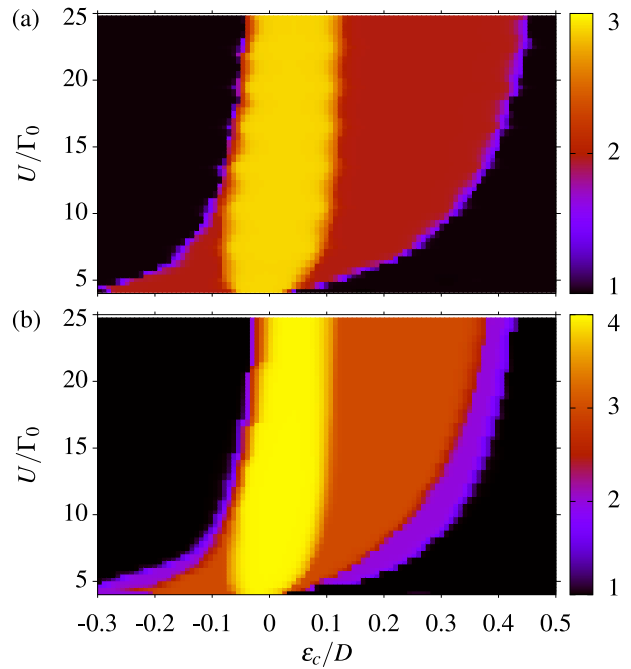}

\caption{$\text{exp}(S_\text{imp}/k_\text{B})$ phase diagram for (a) $N_f=3$ and (b) $N_f=4$ impurities, separated by $\Delta R=2a$ and plotted against $\e_c/D$ and $U/\Gamma_0$ for a constant temperature $T/\Gamma_0=10^{-15}$, $\e^f_l/\Gamma_0=-3$ and $D/\Gamma_0=10$.
 }

\label{fig:FMQCP_mu-U}
\end{center}
\end{figure}

In order to qualitative study this competition between the AF and FM RKKY interaction,
we analyze the depleted Anderson model for a finite number of $N_f=3$ and $N_f=4$ correlated orbitals in 1d, the simplest MIAMs of the second kind. 
However, as already discussed, such QCPs generally occur 
in 
MIAMs of the second kind and, therefore, are not restricted to 1d. 

In Fig.\ \ref{fig:FMQCP_mu-U}, we color plotted the degeneracy  $g=\exp(S_\text{imp}/k_\text{B})$
for the dilute MIAM 
at a fixed temperature $T/\Gamma_0=10^{-15}$ with (a) $N_f=3$ and (b) $N_f=4$ impurities, as a function of the band-filling $\e_c$ and the strength of the Coulomb interaction $U$, with $\epsilon^f/\Gamma_0=-3$ and $D/\Gamma_0=10$.
The phase diagram for $N_f=3$ and $N_f=4$ correlated orbitals is qualitatively identical, separating a spin compensated singlet phase at finite $\e_c$ (black) from a phase with finite degeneracy around $\e_c=0$ (yellow red blue).

The maximal value of the degeneracy near $\e_c=0$ (yellow) is $\text{exp}(S_\text{imp}/k_\text{B})=3$ for $N_f=3$ in Fig.~\ref{fig:FMQCP_mu-U}(a), and $\text{exp}(S_\text{imp}/k_\text{B})=4$ for $N_f=4$ in Fig.~\ref{fig:FMQCP_mu-U}(b). 
At $T=0$, however, this value color coded in yellow can only be reached 
at precisely  $\e_c=0$, where ${\rm rank}(\mat{\Gamma}_\sigma)=1$: for a very small but finite
$|\e_c|>0$, ${\rm rank}(\mat{\Gamma}_\sigma)=2$ such that $g=g_{\rm max}-1$. 
Since one of the two effective hybridizations, $\Gamma_{n\sigma}$, is very small
for small deviations from half-filling, the associated Kondo temperature is exponentially suppressed. Consequently, the extended
yellow region of the maximal degeneracy is only replaced by a region with $g_{\rm max}-1$, once the 
temperature is lower than the exponentially suppressed lowest Kondo temperature.
However, even at $T=0$, two/three regions with different degeneracy of the ground state remain for $N_f=3/N_f=4$. 

As discussed in Sec.~\ref{sec:KT-QPT}, regions with different degeneracy of the ground state are separated by a QCP of Kosterlitz-Thouless type and associated with the decoupling of an effective spin-$1/2$ degree of freedom from the continuum: the coupling of that spin to the continuum changes from antiferromagnetic to ferromagnetic, and the screening breaks down.

The phase boundaries are symmetric with respect to $\e_c$
for PH symmetric impurities ($U=-2\e^f=6\Gamma_0$ in Fig.~\ref{fig:FMQCP_mu-U}).
In the case of $U<-2\e^f$, the degenerate phase is shifted into the $\e_c/D<0$ side of the phase diagram, whereas it is vice versa for $U>-2\e^f$.

In the two-impurity model,  a spin-singlet ground state is always generated for finite $\e_c$
since enough screening channels are available.
The emergence of QCPs due to FM correlations, as reported in this and the previous section, however, 
are a common feature of MIAMs of the second kind.
Since the phase diagram for $N_f=3$ and $N_f=4$, shown in Fig.~\ref{fig:FMQCP_mu-U}, is qualitatively identical for $N_f=5$ (not shown here), and the physical origin of the QCPs is just a consequence of a competition between AF and FM RKKY interactions in combination with the reduced number of available conduction band screening channels, the phase diagram will be qualitatively identical if more correlated orbitals are added.
The geometric arrangement of the correlated orbitals onto the lattice governs the
hybridization matrix $\mat{\Delta}_\sigma(z)$. While the FM QCPs occur in depleted host bands, $|\e_c/D|\to 1$,
in dense impurity setups, they are shifted to half-filling in dilute impurity arrays.
First-order perturbation theory in $U$ revealed that the FM ground state in the depleted model at half-filling also exists in higher spatial dimensions of the lattice \cite{Potthoff2014},
and is not an artifact of a 1d host geometry.
Altogether, these results indicate, that a transition from a delocalized paramagnetic to a localized ferromagnetic state, which is beyond a Hertz–Millis–Moriya spin density wave scenario, might be realizable in the depleted PAM with $\e_c\not=0$.
Due to the exponential suppression of the lowest crossover scale in such models, the 
unstable intermediate fixed points might be more relevant for the experimentally accessible temperatures.

\section{Conclusion and Outlook}
\label{sec:summary}

We presented a  mapping of a strongly correlated multi-impurity models 
onto a correlated cluster model subject to couplings to a number 
of effective conduction bands that is exact in the wide-band limit.
 The AF exchange interaction is encoded
in the cluster orbital hopping matrix, whereas the FM interaction and the competing Kondo effect by the remaining host degrees of freedom are included in the coupling to the effective conduction band channels. This allows us to study the self-screening of the local moments as well as the emerging
local moment fixed points via exact diagonalization of the cluster prior to an investigation of the full problem.
It also opens the door for determining the requirements in the models with potential magnetic
frustrations such as the trimer model. An interesting question arises in particle-hole asymmetric situations beyond the
wide-band limit. Potentially the flow to some asymmetric strong coupling fixed point can increase the number of effective screening channels and will require a more careful analysis.

Since our mapping incorporates the RKKY interaction and the related potential scattering on an equal footing, we were able to show that the frustration induced NFL fixed point of a trimer model with $C_3$ symmetry \cite{PaulIngersent1996,Ingersent2005} is restricted to only a finite range of AF RKKY interaction strength, and can completely disappear in the phase diagram if the Kondo temperature exceeds a certain limit.

For large $N_f$, the MIAMs fall typically into problems of the second kind, where the number of impurities exceeds the number of available Kondo screening channels.  The most prominent example is the PAM.
Since the maximal number of independent conduction bands that couple to the correlated cluster scales with the number of $k$-points on
the Fermi surface, it is limited to $N_b=2$ in 1d. As a consequence, our mapped model predicts a local moment fixed point for the dilute
Anderson lattice in 1d \cite{Titvinidze2015,Potthoff2013} with a large local moment of $I=(N_f-1)/2$.

The physical properties of strongly correlated multi-impurity models strongly depend on whether they are of first or second kind. 
Ferromagnetic correlations between the local moments are irrelevant for the spin-compensated ground state in MIAMs of the first kind, whereas they lead to QCPs of KT type and the suppression of the screening of effective spin-$1/2$ degrees of freedom in models belonging to the second kind.
Therefore, it would be helpful to decide in advance into which kind of category these models fall.

We  demonstrated how the matrix elements of the mapped model depend on the chemical potential or the band center
of the original model. With increasing particle-hole asymmetry, the ground state multiplet in the cluster shows a precursor of a spin-density wave modulation in the spin-spin correlation function. Level crossings between different cluster ground states are associated with magnetic frustrations and an enhancement of the lowest temperature scale of the problem taking the role of the Kondo temperature. The magnetic energy scale, defined via the change of the cluster entropy, however, is decreasing. It vanishes for a decoupled cluster but remains finite due to the asymmetric couplings to the remaining conduction electron bands. 

For an exponential suppression of the Kondo scale, the strong decrease of the magnetic energy scale $T_1$
at the point of magnetic frustration could be misread as a indication of a QCP between different ground states. NRG calculations reveal, however, that there is still a smooth crossover between two different singlet states since the asymmetry is a relevant perturbation for the frustrated isospin Kondo regime  \cite{PaulIngersent1996,Ingersent2005}, just as has been shown for the two-impurity problem \cite{AffleckLudwigJones1995,Silva1996}.  At the band edges the ferromagnetic correlations are recovered even for a small finite-size impurity cluster, and the system shows a series of QCPs of KT type when changing the ground state degeneracy by one.

There is an ongoing debate about a confinement/deconfinement transition in HFs where magnetic order might reduce the number of conduction electrons contributing to the Fermi surface: a heavy FL should be characterized by a large Fermi surface while in a local-moment magnetic metal the correlated electrons are excluded from the Fermi volume \cite{LoehnysenWoelfeReview2007}.
Such a scenario of fractionalized Fermi liquids \cite{SenthilSachdevVojta2003} requires a two-fluid model, 
where the well defined light quasiparticles and a spin-liquid are disconnected. This might be achievable by a Mott-Hubbard insulator transition within the correlated electron subsystem,
in which the remaining spin-spin interactions induce a spin liquid. The lack of charge fluctuation channels suppresses the coherent quasiparticle formation
decoupling the light quasiparticles on a low-energy scale.

Recent experiments \cite{FMQCP_Kotegawa2019,FMQCP_Shen2020,FMQCP_Steppke2013} revealed strange-metal behavior at the FM-PM transition in HFs, and from the theoretical point of view it is known that the FM Kondo lattice possesses a small Fermi surface excluding the local moments \cite{FMQCP_Yamamoto_2010}. Motivated by theses results, we investigated dilute multi-impurity models of the second kind, where the FM RKKY interaction dominates for a roughly half-filled conduction band, and demonstrated the existence of several QCPs of KT type at which local moments decouple from the continuum. This finding indicates a possible confinement/deconfinement transition in a periodic extension of such models, the depleted PAM.

Another fascinating subject is the physics of Kondo holes  \cite{Schlottmann91I,Clare96,Figgins2011,Vojta2014}. Our mapping can provide
some insight into the spatial distribution of the generated local magnetic moments by those holes as well as providing another low-energy scale characterizing the screening of those moments under realistic conditions.


%

\end{document}